\documentclass[a4paper,12pt]{article}
\pdfoutput=1
\usepackage{amsmath}
\usepackage{amsfonts}
\usepackage{amssymb}
\usepackage{graphicx, rotating}
\usepackage{epsfig}
\usepackage{latexsym}
\usepackage{graphicx}
\usepackage{color}
\usepackage{amsmath,bm,amssymb}
\usepackage{color}
\usepackage[english]{babel}
\usepackage{hyperref}
\usepackage[numbers,sort&compress]{natbib}

\newcommand{\Slash}[1]{{\ooalign{\hfil#1\hfil\crcr\raise.167ex\hbox{/}}}}

\newcommand{\beq}{\begin{equation}}  \newcommand{\eeq}{\end{equation}}
\newcommand{\bef}{\begin{figure}}  \newcommand{\eef}{\end{figure}}
\newcommand{\bec}{\begin{center}}  \newcommand{\eec}{\end{center}}
  
\newcommand{\laq}[1]{\label{eq:#1}}

\newcommand{\eq}[1]{(\ref{eq:#1})}

\def\({\left(}
\def\){\right)}

\def\f{\phi}

\def\*{\dagger}

\begin{document}
\renewcommand\bibname{\Large References}

\begin{center}

\hfill   TU-1190

\vspace{1.5cm}

{\Large\bf  The QCD Axion: A Unique Player in the Axiverse with Mixings}

\vspace{1.5cm}

{\bf  Kai Murai, Fuminobu Takahashi, Wen Yin }

\vspace{12pt}
\vspace{1.5cm}
{\em 
Department of Physics, Tohoku University,  
Sendai, Miyagi 980-8578, Japan 
\vspace{5pt}}

\vspace{1.5cm}
\abstract{
In an axiverse with numerous axions, the cosmological moduli problem poses a significant challenge because the abundance of axions can easily exceed that of dark matter.
The well-established stochastic axion scenario offers a simple solution, relying on relatively low-scale inflation.
However, axions are typically subject to mixing due to mass and kinetic terms, which can influence the solution using stochastic dynamics.
Focusing on the fact that the QCD axion has a temperature-dependent mass, unlike other axions, 
we investigate the dynamics of the QCD axion and another axion with mixing.
We find that the QCD axion abundance is significantly enhanced and becomes larger than that of the other axion for a certain range of parameters.
This enhancement widens the parameter regions accounting for dark matter.
In addition, we also find a parameter region in which both axions have enhanced abundances of the same order, which result in multi-component dark matter.
}

\end{center}
\clearpage

\setcounter{page}{1}
\setcounter{footnote}{0}

\section{Introduction}

In string theory or M-theory, many axions appear at low energies~\cite{Witten:1984dg, Svrcek:2006yi,Conlon:2006tq,Arvanitaki:2009fg,Acharya:2010zx, Higaki:2011me, Cicoli:2012sz,Demirtas:2018akl,Marsh:2019bjr, Mehta:2020kwu,Chen:2021hfq,Mehta:2021pwf,Cicoli:2022fzy}. 
Since axion masses are produced by non-perturbative effects, they are expected to span a very wide range. 
A universe with such a large number of axions is called an axiverse~\cite{Arvanitaki:2009fg}. 
One or more of these axions are likely to interact with Standard Model (SM) gauge bosons, and some linear combination of them could be a QCD axion~\cite{Peccei:1977hh, Peccei:1977ur, Weinberg:1977ma,Wilczek:1977pj}.
Interestingly, the presence of many light axions can solve the quality problem of the Peccei-Quinn symmetry of the QCD axion~\cite{Acharya:2010zx}. 
Furthermore, if many ultra-light axions are coupled to photons, this could explain the isotropic cosmic birefringence~\cite{Carroll:1998zi,Finelli:2008jv,Arvanitaki:2009fg,Panda:2010uq,Fedderke:2019ajk,Fujita:2020aqt,Fujita:2020ecn,Takahashi:2020tqv,Mehta:2021pwf,Nakagawa:2021nme,Choi:2021aze,Obata:2021nql,Yin:2021kmx,Kitajima:2022jzz, Gasparotto:2022uqo,Lin:2022niw,Murai:2022zur,Gonzalez:2022mcx,Galaverni:2023zhv}, which is suggested by recent analyses~\cite{Minami:2020odp,Diego-Palazuelos:2022dsq,Eskilt:2022wav,Eskilt:2022cff}. 
In the axiverse, the decay constants of axions are usually thought to be of order the string scale $\sim 10^{15-17}$\,GeV. 
However, it could be much smaller in a setting such as a large volume scenario~\cite{Balasubramanian:2005zx,Conlon:2005ki}.

In the axiverse scenarios, it is known that the abundance of axions produced by the misalignment mechanism~\cite{Preskill:1982cy,Abbott:1982af,Dine:1982ah} will be too large unless  the initial angles are fine-tuned.
This is nothing but the cosmological moduli problem~\cite{deCarlos:1993wie,Banks:1993en}.
In particular, the larger the mass of the axion, the greater the abundance. For example, the QCD axion is known to have excessive abundance when the decay constant is larger than $10^{12}$\,GeV, which sets the upper end of the so-called axion window.

The cosmological moduli problem can be mitigated if the Hubble parameter during inflation is sufficiently small and the inflation lasts long enough.
For example, eternal inflation with a low energy scale fits this scenario~\cite{Kitajima:2019ibn}. 
This is because the initial misalignment angle $\theta_i$ becomes much smaller than ${\cal O}(1)$ if the inflation lasts long enough, allowing the axion field to follow the so-called Bunch-Davies distribution, which balances quantum fluctuations and classical motion.
It has been found that the QCD axion with a decay constant $f_a \gtrsim 10^{12}$\,GeV, which exceeds the upper limit of the axion window, 
is allowed without overclosing the universe, as long as the Hubble parameter is below the QCD scale during inflation~\cite{Graham:2018jyp, Takahashi:2018tdu}.
In Ref.~\cite{Ho:2019ayl}, this stochastic approach was applied for the first time to the cosmological moduli problem posed by numerous axions appearing in the axiverse, not just the QCD axion. 
In this case, the axion abundance is greater for the lighter axion, because although the energy density at the onset of oscillations is the same, the lighter axion starts oscillating later.
Their results show that there is an upper bound on the Hubble parameter during inflation, $H_{\rm inf} < $ {\rm keV - MeV}, depending on the typical decay constant of axions, and that the cosmological moduli problem in the axiverse is solved when this bound is satisfied. 
This was also subsequently confirmed in Ref.~\cite{Reig:2021ipa}. 

Note that the stochastic dynamics can be altered, for example, when a Hubble-induced mass is present~\cite{Alonso-Alvarez:2019ixv}, when the potential deviates from the quadratic potential~\cite{Daido:2016tsj,Nakagawa:2020eeg}, or when the QCD gauge coupling is strong during inflation~\cite{Matsui:2020wfx}.
In such cases, the axion abundances are known to change.
In particular, an important implicit assumption in stochastic axion scenarios is that the axion minima do not change during and after inflation~\cite{Takahashi:2018tdu}. 
This assumption is violated, e.g., if the inflaton is an axion that mixes with other light axions.
In such cases, the probability distribution of axions could shift from the potential minimum to near the maximum, leading to a significant increase in the abundance of axions.
Consequently, it becomes possible to explain all dark matter with axions with the decay constants as small as the astrophysical lower bound~\cite{Takahashi:2019qmh, Takahashi:2019pqf, Nakagawa:2020eeg}. 

In addition, most analyses of axions in the axiverse so far have not considered mixing between axions, and cosmological and astrophysical effects have been studied for individual axions because the axion masses are hierarchical.
On the other hand, it has been pointed out in Refs.~\cite{Kim:2004rp,Choi:2014rja,Higaki:2014mwa,Kaplan:2015fuy,Giudice:2016yja,Higaki:2016jjh,Higaki:2016yqk,Gavela:2023tzu} that the mixing of many axions through mass and kinetic terms can have significant cosmological consequences, and a typical example is  the so-called clockwork/alignment mechanism.
Another interesting phenomenon that is characteristic when the QCD axion is composed of multiple axions is resonance phenomena similar to the MSW effect in neutrino oscillations~\cite{Kitajima:2014xla,Daido:2015cba,Ho:2018qur}.
Through the resonance, the QCD axion can be converted to lighter axion-like particles and vice versa. 
In some cases, the axion starts to run along a lighter flat direction, going over potential hills and troughs~\cite{Daido:2015bva}. 
When multiple axions are present and mixed with each other in this way, the dynamics can lead to complex and interesting phenomena.

In this paper, we study for the first time the mixing effect between string axions and QCD axion under stochastic dynamics.
Normally, in stochastic axion scenarios, the initial field values are determined by the equilibrium distributions during inflation.
However, in the presence of mixing effects and temperature dependence of the mass, we find that the initial values set during inflation can be significantly modified by the post-inflationary axion dynamics. 
Specifically, when the axion potential is generated by QCD instanton effects and another non-perturbative effect, the axions mix through the mass term, causing the mass eigenstates to vary in time due to the temperature dependence of the QCD potential. 
We find that if each of the two potentials has a mass of the same order at the onset of field oscillations, the total energy density of the axions can be significantly enhanced compared to the case without mixing.
In particular, while the lighter axion tends to have a larger abundance in the stochastic scenario, the QCD axion, which is heavier than the mixing partner in the vacuum, can dominate the abundance due to this enhancement.
This enhancement breaks the one-to-one correspondence between the axion mass and the abundance, thus broadening the parameter range that explains the dark matter.
This effect is analogous to the aforementioned shift of the potential minimum due to mixing with the inflaton, but in our scenario, it involves only simple dynamics of axions.
Also, unlike resonance phenomena, there need be no adiabatic invariants and therefore no large mass hierarchy.
This is therefore an example of how mixing between axions and temperature dependence can be very important in axiverse scenarios, especially with stochastic axions.

The rest of this paper is organized as follows.
In Sec.~\ref{sec: model}, we show the model and explain the dynamics of the axions during and after inflation.
In Sec.~\ref{sec: enhancement}, we show the results of the numerical simulations of the axion dynamics and demonstrate the enhancement of the axion energy density.
Finally, we summarize and discuss the results in Sec.~\ref{sec: summary}.

\section{Stochastic axions with mixings}
\label{sec: model}

\subsection{Set-up}

In the string axiverse, there exist numerous axions at low energies, which acquire potentials from non-perturbative effects such as strong dynamics in hidden gauge sectors.
Additionally, compactification of extra dimensions typically induces small instanton effects that generate potentials for axions. 
As a result of these potential terms, axions generically get mixed with each other.
To solve the strong CP problem using these axions, we need axions coupled to gluons, and at least one of them must be extremely light when we switch off the non-perturbative QCD effects.
The requirement for such light axions coupled to gluons is nothing more than the quality problem of the Peccei-Quinn symmetry.
This problem can be solved naturally in the axiverse~\cite{Acharya:2010zx}, where there are many axions whose mass is very light and spans a very wide range.

Although many axions may exist over a wide range of scales, for our interest in the mixing between the QCD axion and other axions, it is sufficient to consider two axions whose masses are not too far apart. 
We introduce two axions, $a$ and $\phi$, and identify their linear combination as the QCD axion, as described below.
We consider a low-energy effective Lagrangian for $a$ and $\phi$ given by
\begin{align}
    \mathcal{L}
    =
    \frac{1}{2} \partial_\mu a \partial^\mu a
    + \frac{1}{2} \partial_\mu \phi \partial^\mu \phi
    - V(a, \phi)
    \ ,
\end{align}
with the potential of%
\footnote{
We assume that the potential $V_{a\phi}$ is time-independent. 
For instance, this is the case if it comes from some non-perturbative effects in a hidden sector, whose typical energy scale is always lower than the dynamical scale. 
This may require that the inflaton primarily reheat the SM sector and not the hidden sector.
See Refs.~\cite{Arias:2012az,Nakagawa:2022wwm} for the case that the dark sector is reheated and the issue in cooling the dark sector~\cite{Nakagawa:2022wwm}.}
\begin{align}
    V(a, \phi)
    &=
    V_\mathrm{QCD}(a) + V_{a \phi}(a,\phi)
    \nonumber \\
    &=
    \chi(T) \left[ 
        1 - \cos \left( \frac{a}{f_a} \right)    
    \right]
    +
    m_\phi^2 f_\phi^2 \left[ 
        1 - \cos \left( N\frac{a}{f_a} + \frac{\phi}{f_\phi} \right)    
    \right]
    .
\end{align}
Here $\chi(T)$ denotes the topological susceptibility of QCD, 
\begin{align}
    \chi(T)
    =
    \left\{
        \begin{array}{ll}
            \chi_0 & \quad (T < T_\mathrm{QCD})
            \\
            \chi_0 \left( \frac{T}{T_\mathrm{QCD}} \right)^n & \quad (T \geq T_\mathrm{QCD})
        \end{array}
    \right.
    \ ,
\end{align} 
where we adopt $n = - 8.16$~\cite{Borsanyi:2016ksw}, $\chi_0 = (75.6\,\mathrm{MeV})^4$, and $T_\mathrm{QCD} = 153\,\mathrm{MeV}$.
We have neglected the higher-order QCD contributions from the $a$-meson mixings in the potential since the axion field evolution will turn out to be around the vicinity of the CP-conserving minimum, where the higher-order terms are irrelevant.
Note that we have used a field redefinition without loss of generality such that $a$ is the combination that couples to gluons.
Then, by taking $m_\phi\to 0$, $a$ becomes the QCD axion, while it is a component of the QCD axion with $m_\phi\neq 0$.
However, since we are mainly interested in the case where $m_\phi$ is smaller than the QCD axion mass in the vacuum,
then $a$ is the main component of the QCD axion.
Therefore, we often refer to $a$ as the QCD axion below.
Also, the constant phase in each potential is absorbed into $a$ and $\phi$.
Thus, the minimum of $V_\mathrm{QCD}$, $a = 0$, is the strong CP conserving point.
Throughout the paper, we concentrate on the possibility 
\beq
    \laq{faff}f_a\sim f_\phi \ ,
\eeq
since we consider that they are both the string axions.

We define the temperature-dependent axion mass $m_a(T)$ from $\chi(T) = m_a^2(T) f_a^2$, which gives the zero-temperature mass
\begin{align}
    m_{a0}
    \equiv 
    m_a(T < T_\mathrm{QCD})
    \approx
    5.7 \times 10^{-9}\,\mathrm{eV} 
    \left( \frac{f_a} {10^{15}\,\mathrm{GeV}} \right)^{-1}
    \ .
\end{align}
Here and hereafter, we denote quantities at the present time by subscript 0.
For later use, we define $\Phi$ and $A$ as
\begin{align}
    \begin{pmatrix}
        \Phi \\ A
    \end{pmatrix}
    \equiv
    \frac{1}{\sqrt{f_a^2 + N^2 f_\phi^2}}
    \begin{pmatrix}
        N f_\phi & f_a
        \\
        - f_a & N f_\phi
    \end{pmatrix}
    \begin{pmatrix}
        a \\ \phi
    \end{pmatrix}
    \ .
\end{align}
Then, $V_{a \phi}(a,\phi)$ is a function of $\Phi$ only, and it is flat in the direction of $A$.

\begin{figure}[!t]
    \begin{center}  
        \includegraphics[width=105mm]{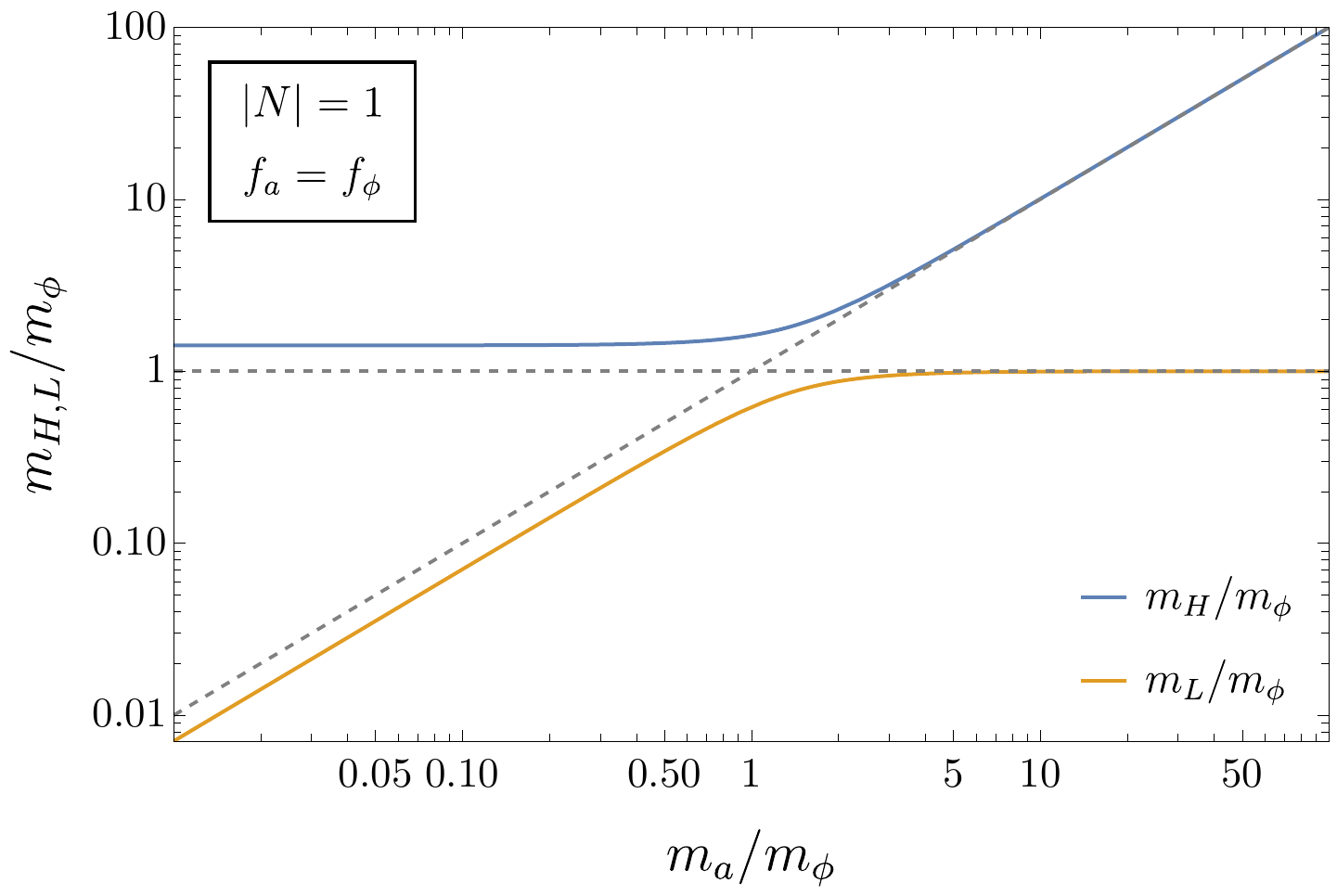}
        \end{center}
    \caption{%
        Dependence of $m_H$ and $m_L$ on $m_a/m_\phi$.
        We set $|N| = 1$ and $f_a = f_\phi$.
        The dashed lines represent $m_{H,L} = m_\phi$ and $m_a$.
    }
    \label{fig: level crossing} 
\end{figure}

Around the origin, $a = \phi = 0$, the potential $V(a, \phi)$ can be approximated by the quadratic terms as
\begin{align}
    V(a, \phi)
    &\simeq
    \frac{1}{2}
    \begin{pmatrix}
        a & \phi
    \end{pmatrix}
    \begin{pmatrix}
        \frac{\chi(T) + N^2 m_\phi^2 f_\phi^2}{f_a^2} & \frac{N m_\phi^2 f_\phi}{f_a}
        \\[0.3em]
        \frac{N m_\phi^2 f_\phi}{f_a} & m_\phi^2
    \end{pmatrix}
    \begin{pmatrix}
        a \\ \phi
    \end{pmatrix}
    \nonumber \\
    &\equiv 
    \frac{1}{2}
    \begin{pmatrix}
        a & \phi
    \end{pmatrix}
    M(T)
    \begin{pmatrix}
        a \\ \phi
    \end{pmatrix}
    \ .
\end{align}
The mass matrix, $M(T)$, is diagonalized by an orthogonal matrix $U$ as
\begin{align}
    U M U^\mathrm{T} 
    =
    \begin{pmatrix}
        m_H^2 & 0
        \\
        0 & m_L^2
    \end{pmatrix}
    \ , \quad
     U
    = 
    \begin{pmatrix}
        \cos \alpha & - \sin \alpha
        \\
        \sin \alpha & \cos \alpha
    \end{pmatrix}
    \ ,
\end{align}
where $\alpha$ is the mixing angle, and we assume $m_{H0} > m_{L0} > 0$ without loss of generality.
Note that both $\alpha$ and $U$ continuously depend on $T$.
The heavy and light mass eigenstates, $s_H$ and $s_L$, are related to $a$ and $\phi$ as
\begin{align}
    \begin{pmatrix}
        s_H \\ s_L
    \end{pmatrix}
    =
    U
    \begin{pmatrix}
        a \\ \phi
    \end{pmatrix}
    \ .
\end{align}
Note that $U$, $m_{H,L}$, and $s_{H,L}$ are all temperature-dependent quantities.
We show the dependences of $m_H$ and $m_L$ on $m_a/m_\phi$ for $f_a = f_\phi$ and $|N| = 1$ in Fig.~\ref{fig: level crossing}.
For $m_a \gg m_\phi$, the mass eigenstates are largely determined by $V_\mathrm{QCD}$ as $s_H \simeq a$ and $s_L \simeq \phi$.
Then, the mass eigenvalues become $m_H \simeq m_a$ and $m_L \simeq m_\phi$.
On the other hand, for $m_a \ll m_\phi$, $V_\mathrm{QCD}$ is negligible, and $\Phi$ and $A$ correspond to $s_H$ and $s_L$, respectively.
We also see that $m_H > m_L$ holds for all $m_a / m_\phi$.
Thus, we refer to $s_H$ and $s_L$ as the heavier and lighter modes regardless of temperature, respectively.
Around $m_a = m_\phi$, the heavier mode transitions from $\Phi$ to $a$ as $m_a / m_\phi$ increases. 
If the adiabatic condition is satisfied during the transition, the resonant conversion between the two axions can take place. 
In the following, however, we do not need the adiabatic condition, and we will see that the axion dynamics is more complicated.

\begin{figure}[!t]
    \begin{center}  
        \includegraphics[width=105mm]{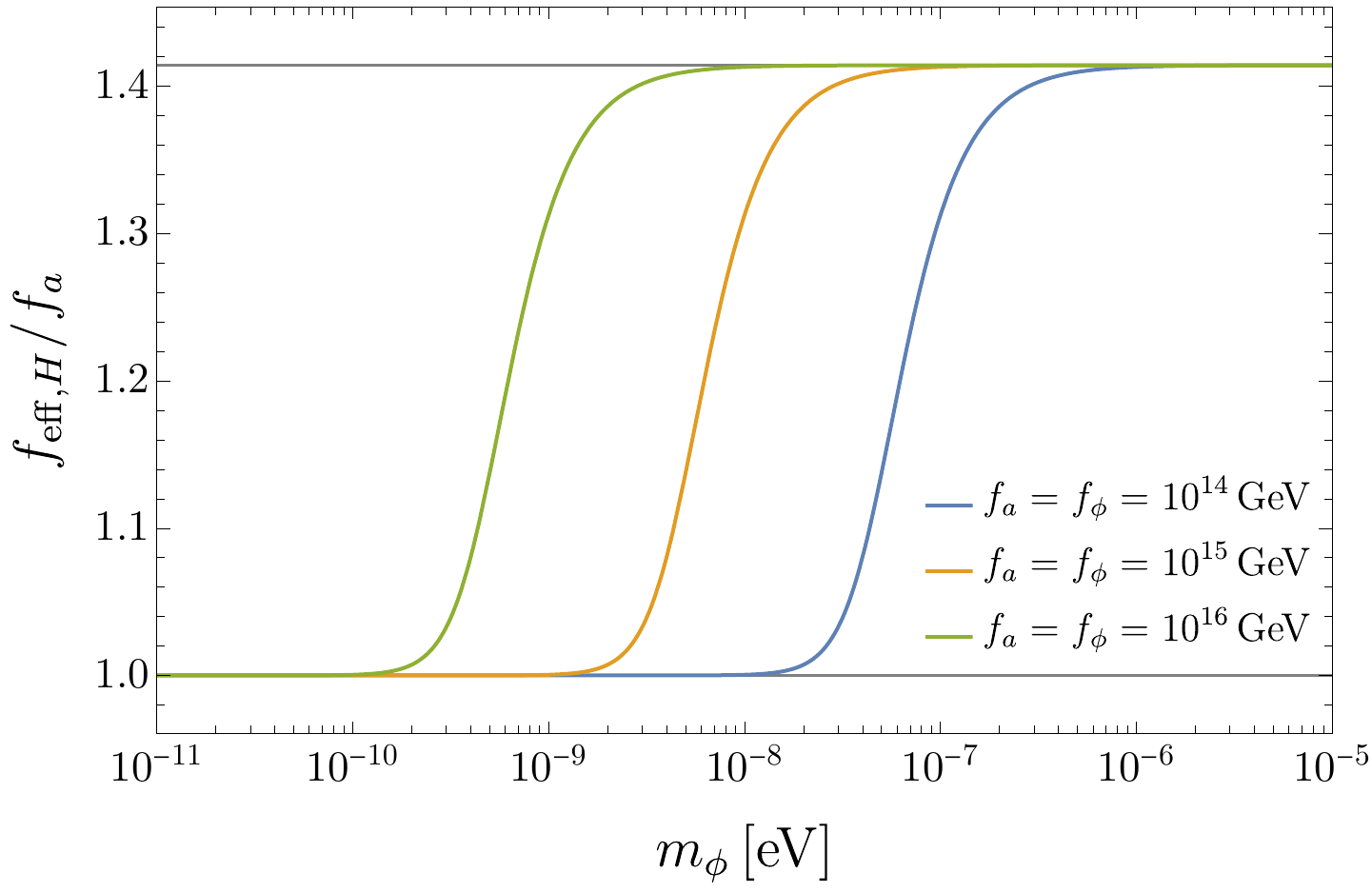}
    \end{center}
    \caption{%
        Dependence of $f_{\mathrm{eff}, H}$ on $m_\phi$. 
        We set $|N| = 1$, $m_a = m_{a 0}$, and $f_a = f_\phi$.
        The gray horizontal lines correspond to $f_{\mathrm{eff},H}/f_a = 1$ and $\sqrt{2}$.
    }
    \label{fig: feff} 
\end{figure}

If $a$ has couplings to the SM particles other than gluons, such as photons, then $s_H$ and $s_L$ will also be coupled to them through the mixing.
From the relation
\begin{align}
    a = U_{1 1} s_H + U_{2 1} s_L
    \ ,
\end{align}
we define the effective decay constant related to $a$ as
\begin{align}
    f_{\mathrm{eff}, H}
    \equiv 
    \frac{f_a}{|U_{1 1}|}
    \ , \quad 
    f_{\mathrm{eff}, L}
    \equiv 
    \frac{f_a}{|U_{2 1}|}
    \ .
    \label{eq: effective decay constants}
\end{align}
Using these quantities, we can interpret that $s_H$ and $s_L$ are coupled to the SM particles with the effective decay constant $f_{\mathrm{eff}, H}$ and $f_{\mathrm{eff}, L}$, respectively.
We show the dependence of $f_{\mathrm{eff}, H}$ on $m_\phi$ for $|N| = 1$, $m_a = m_{a 0}$, and $f_a = f_\phi$ in Fig.~\ref{fig: feff}

\subsection{Stochastic initial conditions set during inflation}
\label{subsec: during inflation}

Here, we discuss the mass eigenstates and their  typical field values during inflation.
During inflation, we assume that the Gibbons-Hawking temperature~\cite{Gibbons:1977mu}, $T_\mathrm{GH} \equiv H_\mathrm{inf}/2\pi$, is much lower than $T_\mathrm{QCD}$.
This assumption implies an upper bound on $H_\mathrm{inf}$ as 
\begin{align}
    H_\mathrm{inf} 
    \ll
    2 \pi T_\mathrm{QCD}
    \simeq 
    0.96\,\mathrm{GeV}
    \ .
\end{align}
In this case, the QCD axion acquires a potential during inflation, and the topological susceptibility is given by $\chi = \chi_0$.
Then, the axion fields are around the potential minimum at $a = \phi = 0$.
Thus, the mass eigenstates are $s_{H0}$ and $s_{L0}$.
If the axion masses, $m_{H0}$ and $m_{L0}$, are much smaller than $H_\mathrm{inf}$, the axion fields diffuse around the origin due to quantum fluctuations.
If the duration of inflation is sufficiently long, the axion field values follow the Bunch-Davies distribution.
In the Bunch-Davies distribution, the variances of the mass eigenstates are given by~\cite{Graham:2018jyp,Takahashi:2018tdu}
\begin{align}
    \sqrt{\langle s_{H0}^2 \rangle}
    =
    \sqrt{\frac{3}{8\pi^2}} \frac{H_\mathrm{inf}^2}{m_{H0}}
    \ , \quad 
    \sqrt{\langle s_{L0}^2 \rangle}
    =
    \sqrt{\frac{3}{8\pi^2}} \frac{H_\mathrm{inf}^2}{m_{L0}}
    \ .
\end{align}
The field values at the end of inflation play a role of the initial condition for the field dynamics after inflation.
In the following, we parameterize $s_{H0}$ and $s_{L0}$ at the end of inflation by
\begin{align}
    s_{H0, \mathrm{init}}
    =
    c_H \sqrt{\frac{3}{8\pi^2}} \frac{H_\mathrm{inf}^2}{m_{H0}}
    \ , \quad 
    s_{L0, \mathrm{init}}
    =
    c_L \sqrt{\frac{3}{8\pi^2}} \frac{H_\mathrm{inf}^2}{m_{L0}}
    \ ,
\end{align}
where $c_H$ and $c_L$ are typically of $\mathcal{O}(1)$.
Then, the initial values of $a$ and $\phi$ are given by
\begin{align}
    \begin{pmatrix}
        a_\mathrm{init} \\ \phi_\mathrm{init}
    \end{pmatrix}
    =
    U_0^\mathrm{T}
    \begin{pmatrix}
        s_{H0,\mathrm{init}} \\ s_{L0,\mathrm{init}}
    \end{pmatrix}
    \ .
\end{align}

Let us consider two limiting cases: $m_{a0} \gg m_\phi$ and $m_{a0} \ll m_\phi$ with no hierarchy between $f_a$ and $N f_\phi$.
First, in the limit of $m_{a0} \gg m_\phi$, the potential is dominated by $V_\mathrm{QCD}$, and the mass eigenvalues become
\begin{align}
    m_{H0}^2 \simeq m_{a0}^2
    \ , \quad  
    m_{L0}^2 \simeq m_\phi^2
    \ ,
\end{align}
during inflation.
The matrix $U_0$ is given by
\begin{align}
    U
    = 
    \begin{pmatrix}
        \cos \alpha & - \sin \alpha
        \\
        \sin \alpha & \cos \alpha
    \end{pmatrix}
    \ \, {\rm with}
   ~~
    \tan \alpha
    \simeq
    - N\frac{m_\phi^2 f_\phi}{m_{a0}^2 f_a}
    \ ,
\end{align}
where one can see that the mixing angle, $\alpha$, is
very small with $f_a\sim f_\phi$.
In this case, the initial condition is approximated by
\begin{align}
\label{ainit}
    a_\mathrm{init}
    &\simeq
    c_H \sqrt{\frac{3}{8\pi^2}} \frac{H_\mathrm{inf}^2}{m_{a0}}
    \ ,
    \\
    \phi_\mathrm{init}
    &\simeq
    c_L \sqrt{\frac{3}{8\pi^2}} \frac{H_\mathrm{inf}^2}{m_\phi}
    \ .
    \label{phiinit}
\end{align}
Note that we have $|a_\mathrm{init}| \ll  |\phi_\mathrm{init}|$ for $c_H, c_L= {\cal O}(1)$ in this case.

Next, we consider the limit of $m_{a 0} \ll m_\phi$.
In this limit, the potential is dominated by $V_{a \phi}$, and the mass eigenvalues become
\begin{align}
    m_{H0}^2
    \simeq
    m_\Phi^2 
    \equiv 
    \frac{f_a^2 + N^2 f_\phi^2}{f_a^2} m_\phi^2
    \ , \quad  
    m_{L0}^2
    \simeq
    m_A^2
    \equiv 
    \frac{f_a^2}{f_a^2 + N^2 f_\phi^2} m_{a0}^2
    \ .
\end{align}
The matrix $U$ is given by
\begin{align}
    U_0
    =
    \begin{pmatrix}
        \cos \alpha & - \sin \alpha
        \\
        \sin \alpha & \cos \alpha
    \end{pmatrix}
    \ \, {\rm with}
    ~~
    \tan \alpha
    \simeq
    - \frac{f_a}{N f_\phi}
    \ .
\end{align}
Thus, the mass eigenstates are given by
\begin{align}
    s_H \simeq \Phi
    \ , \quad 
    s_L \simeq A
    \ ,
\end{align}
and the initial conditions become 
\begin{align}
    a_\mathrm{init}
    &=
    \sqrt{\frac{3}{8\pi^2(f_a^2 + N^2 f_\phi^2)}} H_\mathrm{inf}^2
    \left( \frac{c_H N f_\phi}{m_{H0}} - \frac{c_L f_a}{m_{L0}} \right)
    \ ,
    \\
    \phi_\mathrm{init}
    &=
    \sqrt{\frac{3}{8\pi^2(f_a^2 + N^2 f_\phi^2)}} H_\mathrm{inf}^2
    \left( \frac{c_H f_a}{m_{H0}} + \frac{c_L N f_\phi}{m_{L0}} \right)
    \ .
\end{align}
In this case, $\Phi$ approximately remains the mass eigenstate during and after inflation, and the dynamics of the two mass eigenstates always decouple.
So in the following we will focus on the case of $m_{a0} \gtrsim m_\phi $ to see how the two fields evolve via the mixing.
Moreover, if $m_\phi \ll H(T_\mathrm{QCD})$, with $H(T)$ being the Hubble parameter at the cosmic temperature $T$ in the radiation dominated Universe, the QCD axion begins to oscillate first, and the other axion begins to oscillate after $m_a(T)$ becomes constant.
In this case, the field dynamics are again independent for each of the two axions.
Thus, we  focus on  $m_\phi$ in the mass range of 
\begin{align}
    3H(T_\mathrm{QCD}) \lesssim m_\phi \lesssim m_{a0}
    \ ,
\end{align}
in the following.

Here, we make two comments on the assumptions of our analysis.
First, to obtain the Bunch-Davies distributions for the axions, the duration of inflation should be sufficiently long as 
\begin{align}
    N_\mathrm{inf} > \max_i{N_{\rm rela}^{i}} 
    \ ,
\end{align}
with $N_\mathrm{inf}$ being the e-folding number of the inflationary period and $N_{\rm rela}^i=H_\mathrm{inf}^2/m_{i}^2$ being the e-folding number required for relaxation of the $i$-th axion with the mass eigenvalue $m_i$. 
With multiple axions, the lightest one requires the longest relaxation time, and thus we evaluate the e-folding number for the lightest mass eigenvalue in our setup.  
For example, this condition becomes $N_\mathrm{inf} >10^{32}$ for $H_\mathrm{inf} = 5$\,MeV and $m_{L0} = 5 \times 10^{-10}$\,eV.
Such long inflation does not necessarily require eternal inflation%
\footnote{
    Note that the typical number of e-folds of eternal inflation is finite~\cite{Barenboim:2016mmw,Assadullahi:2016gkk}, and the eternity of eternal inflation relies on the volume measure. In fact,
    one can make the typical number of e-folds extremely large in a certain model of stochastic inflation~\cite{Kitajima:2019ibn}.
}
since the upper bound to have non-eternal inflation is $N_\mathrm{inf} < 2\pi^2 M_{\rm Pl}^2/3H_\mathrm{inf}^2 \sim 2\times 10^{42}$~\cite{Dubovsky:2011uy, Graham:2018jyp}.

Second, the fluctuations of the axion fields can spatially modulate the Hubble parameter during inflation.
This can be  checked by using the aforementioned relaxation time scale, $N_{\rm rela}^i.$
Without any mixing effect and assuming the quadratic potential, we can consider the constraint for each axion potential, which contributes to the expansion via $\Delta_i H_\mathrm{inf} = \frac{m_{i}^2 f_i^2 }{6M^2_{\rm Pl}H_\mathrm{inf}},$ where $m_{i}^2 f_i^2 $ represents a typical scale of the axion potential height.
The backreaction can be neglected if $\Delta_i H_\mathrm{inf} \times N^i_{\rm rela}/H_\mathrm{inf}\ll 1,$ leading to $f_i\ll M_{\rm Pl}/\sqrt{6}.$
This is consistent with the one from Ref.~\cite{Graham:2018jyp}.

In the presence of mixings or generic potential shapes, we need to solve the Fokker-Planck equation~\cite{Starobinsky:1986fx,Starobinsky:1994bd} with the volume effect~\cite{Nakao:1988yi, Nambu:1988je, Nambu:1989uf, Linde:1993xx}. 
Here, we analytically derive a conservative bound. With multiple axions, the inflationary Hubble parameter is altered at most by 
\begin{equation} 
    \Delta H_\mathrm{inf}
    \approx \frac{\Lambda_{\rm tot}^4}{6M^2_{\rm Pl}H_\mathrm{inf}} 
    \ ,
\end{equation}
where $\Lambda_{\rm tot}^4$ is the total potential height for the multiple axions. 
If $\Delta H_\mathrm{inf}/H_{\rm inf}\times \max_i N^{i}_{\rm rela} \ll 1$ is satisfied, we need not worry about the back reaction.
For our system, we obtain $\Lambda_{\rm tot}^4= 2(\chi_0+ m_\phi^2 f_\phi^2)\simeq 2 \chi_0,$ and $\Delta H_\mathrm{inf} \approx 7\times 10^{-38} H_\mathrm{inf}$.
Thus, $\Delta H_\mathrm{inf}/H_{\rm inf} \times \max_i{N_{\rm rela}^{i}}\sim 10^{-5}$ and we certainly have a parameter region that we can safely neglect the backreaction.
In particular, we can  neglect the backreaction effect in the whole range of parameters shown in Fig.~\ref{fig: Hubble for DM f14}.

\subsection{Post-inflationary dynamics before QCD phase transition}
\label{subsec: RD}

Next, we consider the field dynamics after inflation. 
After the end of the inflation, the inflaton decays into the SM particles, which form a hot thermal plasma. 
As the temperature of the universe increases and becomes higher than the QCD scale, the axion potential from non-perturbative QCD effects, $V_{\rm QCD}$, disappears, and we have only the potential as
\begin{align}
    V(a, \phi)
    \simeq
    m_\phi^2 f_\phi^2 \left[ 
        1 - \cos \left( N\frac{a}{f_a} + \frac{\phi}{f_\phi} \right)    
    \right]
    \ ,
\end{align}
which is only a function of  $\Phi$. 
Depending on the relative size of $m_\Phi$ (or $m_\phi$) and $m_a$, the evolution of the system is different.
When $m_\Phi \simeq 3 H(T) > 3 H(T_{\rm QCD})$, $\Phi$ starts to evolve while $A$ remains constant.
After that, the field dynamics depends on when $V_\mathrm{QCD}$ becomes relevant.

For example, let us consider the limit of $m_\Phi \gg H(T_\mathrm{QCD})$ and $m_{a 0} \gg m_\phi$.
In this limit, $\Phi$ damps due to oscillations well before $T=T_\mathrm{QCD}$.
Thus, the axions settle down at the potential minimum of $V_{a \phi}$, $(a_m, \phi_m)$, determined by
\begin{align}
    N \frac{a_m}{f_a} + \frac{\phi_m}{f_\phi}
    & = 0 \ ,
    \\
    - \frac{a_m}{f_\phi} + N \frac{\phi_m}{f_a}
    &=
    - \frac{a_\mathrm{init}}{f_\phi} + N \frac{\phi_\mathrm{init}}{f_a}
    \ .
\end{align}
As a result, we obtain
\begin{align}
    a_m 
    &=
    \frac{f_a^2 a_\mathrm{init} - N f_a f_\phi \phi_\mathrm{init}}{f_a^2 + N^2 f_\phi^2}
    =
    \frac{m_\phi f_a - \frac{c_L}{c_H} N m_{a0} f_\phi}{f_a^2 + N^2 f_\phi^2}
    \frac{f_a}{m_\phi} a_\mathrm{init}
    \ ,
    \label{eq: a before QCD}
    \\
    \phi_m
    &= 
    \frac{- N f_a f_\phi a_\mathrm{init} + N^2 f_\phi^2 \phi_\mathrm{init}}{f_a^2 + N^2 f_\phi^2}
    =
    - \frac{\frac{c_H}{c_L} m_\phi f_a - N m_{a0} f_\phi}{f_a^2 + N^2 f_\phi^2}
    \frac{N f_\phi}{m_{a0}} \phi_\mathrm{init}
    \ ,
\end{align}
where we have used Eqs.~(\ref{ainit}) and (\ref{phiinit}) in the second equalities.
Note that, since we are assuming $m_{a0} \gg m_\phi$, we have $a_\mathrm{init} \ll \phi_\mathrm{init}$.
Thus, for $f_a \sim |N| f_\phi$,%
\footnote{%
We also note for $m_{a0}\gg m_\f$ with $c_H\sim c_L$ that $f_a\sim |N| f_\phi$ maximizes $|a_m/a_{\rm ini}|$, and thus we cannot get a much further enhancement by relaxing the condition~\eq{faff}.} 
the field values of the axions are modified by the post-inflationary dynamics as
\begin{align}
    a_m &
    =
    \mathcal{O}(1) \phi_\mathrm{init}
    =
    \mathcal{O} \left( \frac{c_L m_{a0}}{c_H m_\phi} \right) a_\mathrm{init}
    \gg a_\mathrm{init}
    \ ,
    \\
    \phi_m
    &
    =
    \mathcal{O}(1) \phi_\mathrm{init}
    \ .
\end{align}
We note that $a_m$ is enhanced compared to $a_\mathrm{init}$, which is crucial for the evaluation of the axion abundances, as we will see in the next section. 
Since $a_m / a_\mathrm{init}$ is proportional to $m_\phi^{-1}$, the enhancement will be more significant for smaller $m_\phi$ as long as $\Phi$ starts to oscillate before $V_\mathrm{QCD}$ becomes relevant.
Thus, we expect that the enhancement is most significant for $m_\phi$ such that $V_\mathrm{QCD}$ and $V_{a \phi}$ becomes relevant at about the same time.
In other words, we expect the enhancement for $m_\phi \sim m_\mathrm{enh} \equiv 3 H(T_{a,\mathrm{osc}})$ with $T_{a,\mathrm{osc}}$ satisfying $m_a(T_{a,\mathrm{osc}}) = 3 H(T_{a,\mathrm{osc}})$.
Considering $H(T) \propto T^2$ and $m_a(T) \propto T^{-4.08}$, we obtain $a_m / a_\mathrm{init} \propto f_a^{-2.04/3.04}$ for $m_\phi \sim m_\mathrm{enh}$, which leads to the maximal enhancement of the QCD axion abundance with a factor proportional to $f_a^{-4.08/3.04} \simeq f_a^{-1.34}$.
Note that here we neglected the temperature dependence of the effective degrees of freedom of radiation, $g_*$, in the Hubble parameter. 

\section{Enhancement of the QCD axion abundance}
\label{sec: enhancement}

In the previous section, we have seen that the amplitude of the QCD axion becomes larger than the initial value due to post-inflationary dynamics caused by mixing. Numerical calculations are needed to determine when and to what extent the QCD axion abundance indeed increases.

\subsection{Setup for numerical calculations}

Here, we perform the numerical calculation of the axion dynamics and show how the enhancement of the axion abundance depends on the model parameters.

The equations of motion for $a$ and $\phi$ are given by
\begin{align}
    \ddot{a} + 3 H \dot{a} + \frac{\partial V(a, \phi)}{\partial a}
    =
    0
    \ , 
    \nonumber \\
    \ddot{\phi} + 3 H \dot{\phi} + \frac{\partial V(a, \phi)}{\partial \phi}
    =
    0
    \ ,
\end{align}
where the dots represent derivatives with respect to the physical time $t$, and the Hubble parameter, $H$, is given by the cosmic temperature $T$ through the Friedmann equation in the radiation-dominated era:
\begin{align}
    3 M_\mathrm{Pl}^2 H^2(T)
    =
    \frac{\pi^2}{30} g_*(T) T^4
    \ .
\end{align}
The time evolution of the temperature is determined by the conservation of the entropy in the physical volume $\propto R^3$ with the scale factor $R$:
\begin{align}
    g_{*s}(T) T^3 R^3 
    =
    \mathrm{const.}
    \ ,
\end{align}
which leads to
\begin{align}
    \dot{T}
    =
    - \sqrt{\frac{\pi^2}{10} g_{*s}(T)} \frac{T^2}{M_\mathrm{Pl}}
    \left( \frac{1}{g_{*s}(T)} \frac{\mathrm{d} g_{*s}(T)}{\mathrm{d} T} + \frac{3}{T} \right)^{-1}
    \ .
\end{align}
We use the temperature dependence of the effective degrees of freedom of radiation for energy density and entropy density, $g_*(T)$ and $g_{*s}(T)$, given in Ref.~\cite{Borsanyi:2016ksw}.

We set the initial conditions
\begin{align}
    a(t=0) = a_\mathrm{init}
    \ , \quad 
    \dot{a}(t=0) = 0
    \ , \quad 
    \phi(t=0) = \phi_\mathrm{init}
    \ , \quad 
    \dot{\phi}(t=0) = 0
    \ , \quad 
    T(t=0) = T_\mathrm{init}
    \ , \quad 
\end{align}
with the initial temperature $T_\mathrm{init}$ satisfying
\begin{align}
    3 M_\mathrm{Pl}^2 (100 m_\phi)^2 
    =
    \frac{61.75 \pi^2}{30} T_\mathrm{init}^4
    \ ,
\end{align}
which corresponds to a time well before the onset of oscillations of $\phi$.
The final time of the simulations is set to be well after the energy densities of $s_H$ and $s_L$ come to follow $\propto R^{-3}$.

The energy density can be expressed in terms of the current density parameter as
\begin{align}
    \Omega_\mathrm{mix} 
    =
    \frac{\rho_{\mathrm{mix}, 0}}{\rho_{c}}
    \ ,
\end{align}
where $\rho_\mathrm{mix}$ is the sum of the energy density of $a$ and $\phi$, and $\rho_c$ is the critical density.
Since the energy density of oscillating scalars scales proportionally to the entropy density, we evaluate $\rho_\mathrm{mix}/s$ at the end of the numerical calculations and obtain
\begin{align}
    \Omega_\mathrm{mix} 
    =
    \frac{\rho_{\mathrm{mix}}}{s} 
    \left( \frac{\rho_c}{s_0} \right)^{-1}
    \ ,
\end{align}
where $\rho_c/s_0 \simeq 3.6 \times 10^{-9} h^2$\,GeV with the reduced Hubble constant $h \simeq 0.67$.

The input parameters of the numerical calculations are
\begin{align}
    \left\{ m_{a0} (\mathrm{or}~f_a), m_\phi, f_\phi, N, c_H, c_L, H_\mathrm{inf} \right\}
    \ .
\end{align}
For simplicity, we fix
\begin{align}
    N = -1
    \ , \quad
    f_a = f_\phi = f
    \ , \quad 
    c_H = c_L = 1 
    \ .
\end{align}
Since we are interested in the low-scale inflation and the Bunch-Davies distribution whose width is much smaller than the decay constant, the potential can be approximated by the quadratic terms.
In this case, $|N|$ is degenerate with $f_\phi$, and the sign of $N$ can also be absorbed into the definition of $\phi$.
The dependence of the axion abundances on the initial conditions, $c_H$ and $c_L$, will be discussed later.
Now, the remaining parameters are $f$, $m_\phi$, and $H_\mathrm{inf}$.
As long as the quadratic approximation is valid, the field values always scale as $\propto H_\mathrm{inf}^2$, and thus the choice of $H_\mathrm{inf}$ does not affect the axion dynamics qualitatively.

To see the non-trivial dynamics of axions due to mixing effects, we focus mainly on $3 H(T_\mathrm{QCD}) \lesssim m_\phi \lesssim m_{a0}$ as mentioned above.
From the Friedmann equation, 
\begin{align}
    3 M_\mathrm{Pl}^2 H(T_\mathrm{QCD})^2
    =
    \frac{\pi^2}{30} g_*(T_\mathrm{QCD}) T_\mathrm{QCD}^4
    \ ,
\end{align}
we obtain
\begin{align}
    H(T_\mathrm{QCD})
    \simeq 
    1.7 \times 10^{-11}\,\mathrm{eV}
    \ .
\end{align}
Thus, we will investigate the mass range of $10^{-11}\,\mathrm{eV} \lesssim m_\phi \lesssim m_{a0}$.

\subsection{Numerical results}

\subsubsection{Axion dynamics}

In the following, we show the numerical results for 
\begin{align}
    f
    =
    10^{15}\,\mathrm{GeV}
    \ , \quad 
    H_\mathrm{inf} 
    =
    5\,\mathrm{MeV}
    \ ,
\end{align}
which corresponds to
\begin{align}
    m_{a0} 
    =
    5.7 \times 10^{-9}\,\mathrm{eV}
    \ .
\end{align}
We consider the following three values of $m_\phi$,
\begin{align}
    m_\phi
    =
    10^{-10.5}\,\mathrm{eV}
    \ ,~
    10^{-9.5}\,\mathrm{eV}
    \ ,~\mathrm{and}~
    10^{-8}\,\mathrm{eV}
    \ ,
\end{align}
as examples for the dynamics with enhancement ($m_\phi = 10^{-9.5}$\,eV) and without enhancement ($m_\phi = 10^{-10.5}$\,eV and $10^{-8}$\,eV).

\begin{figure}[!p]
    \begin{center}  
        \includegraphics[width=90mm]{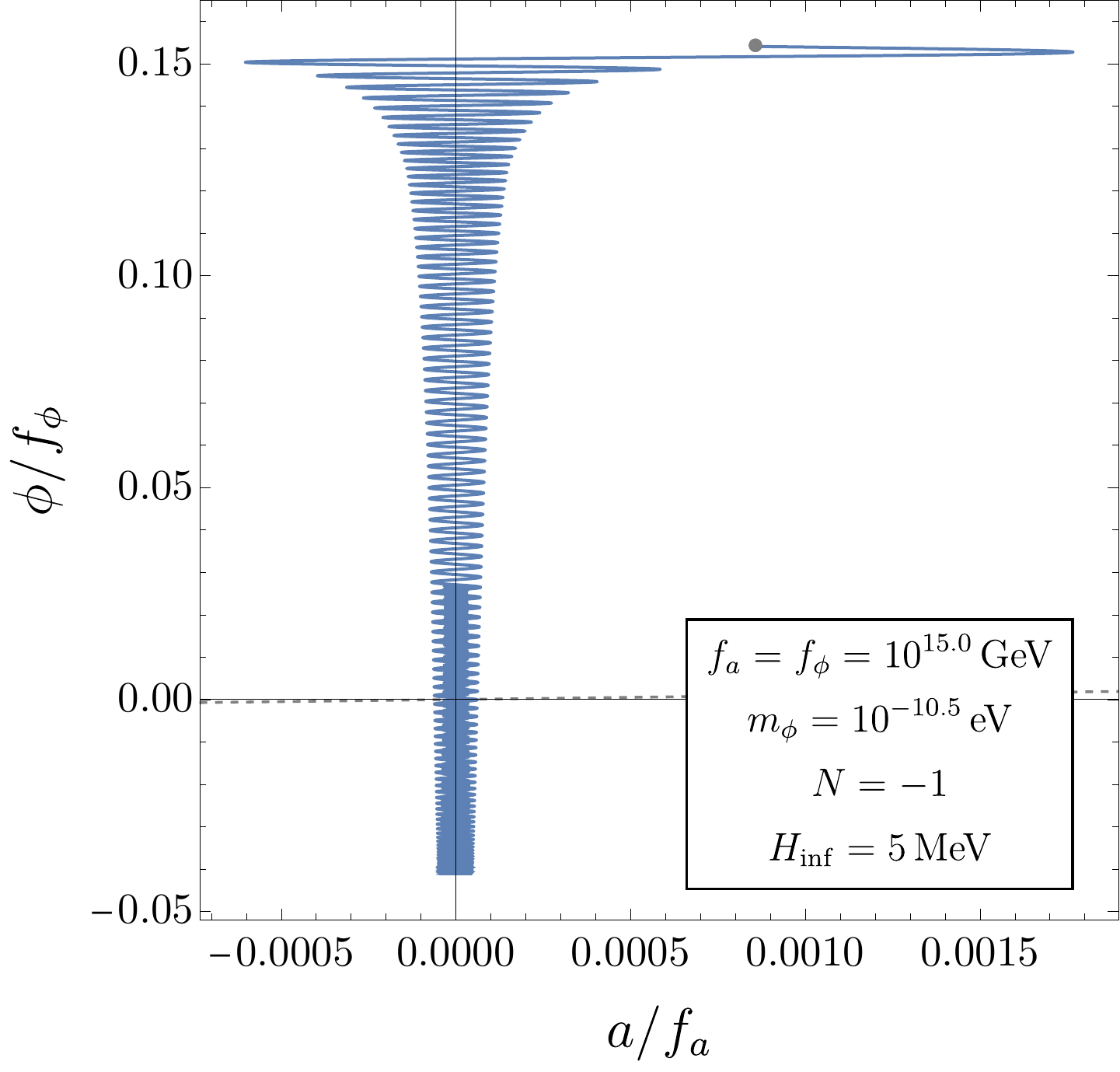}
        \\
        \vspace{5mm}
        \includegraphics[width=110mm]{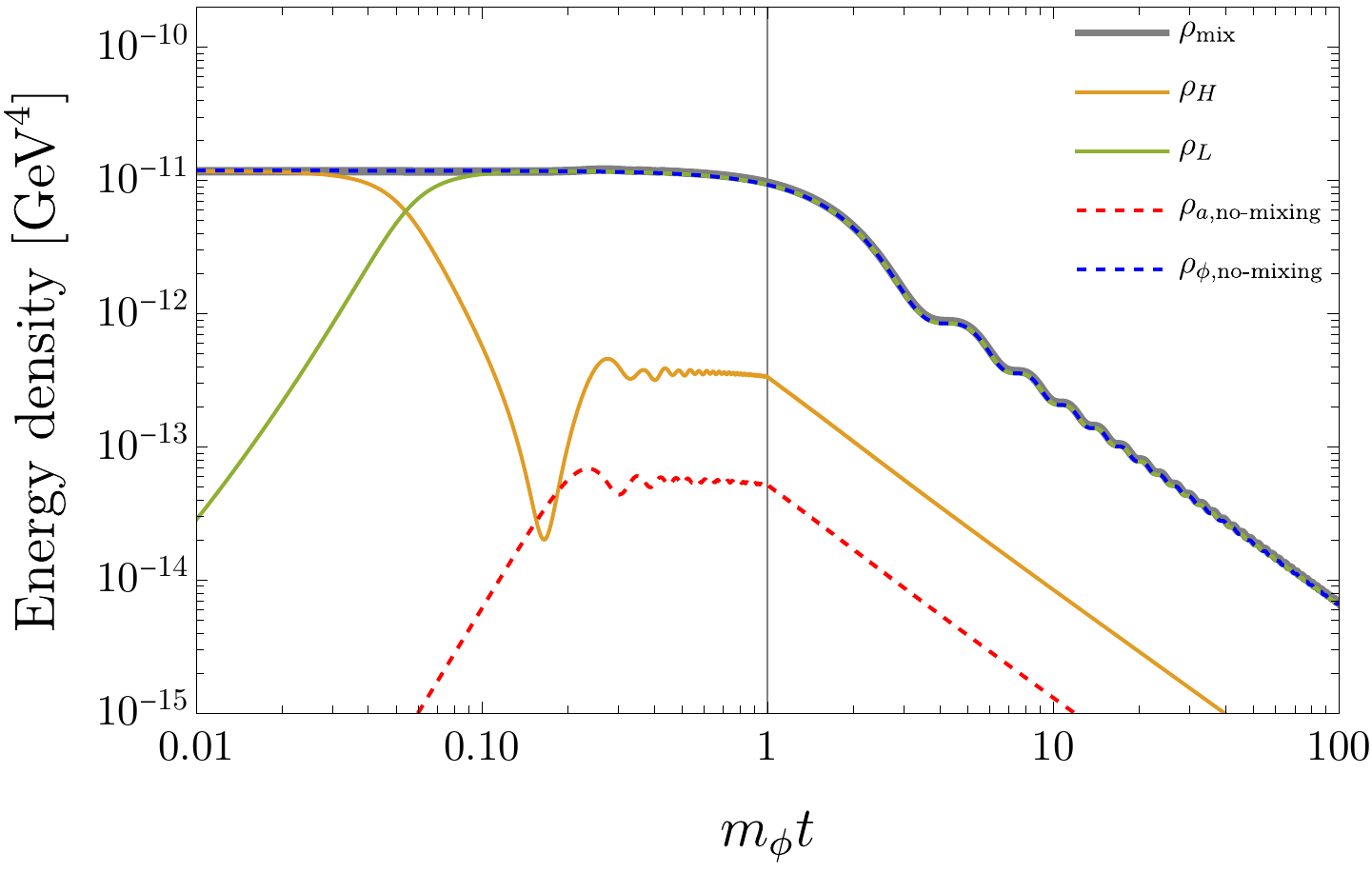}
        \end{center}
    \caption{%
        (Top panel)
        Trajectory of $a/f_a$ and $\phi/f_\phi$.
        The gray dashed line represents $a/f_a = \phi/f_\phi$ (which almost overlaps with $\phi/f_\phi=0$), and the gray dot represents the initial field values.
        (Bottom panel)
        Energy densities of the two fields, heavier mode, and lighter mode with mixing and $a$ and $\phi$ without mixing.
        $\rho_\mathrm{mix}$ almost overlaps $\rho_{\phi,\mathrm{no}\text{-}\mathrm{mixing}}$.
        The vertical line represents $m_\phi t = 1$, which almost corresponds to $T = T_\mathrm{QCD}$.
        As is usually the case with stochastic axions, the lighter axion has a larger abundance.
    }
    \label{fig: evolution -10.5} 
\end{figure}

First, we show the result for $m_\phi = 10^{-10.5}$\,eV in Fig.~\ref{fig: evolution -10.5}.
The top panel shows the trajectory of $a/f_a$ and $\phi/f_\phi$.
Initially, $a/f_a$ is smaller than $\phi/f_\phi$ because of $m_{a0} \gg m_\phi$.
At the very beginning, the axion fields slowly roll down the potential in the $\Phi$-direction.
In the figure, it first moves to the right.
Then, $V_\mathrm{QCD}$ grows and the axion field starts to oscillate rapidly in the $a$-direction.
After that, the fields also start to oscillate in the $\phi$-direction.
The bottom panel shows the time evolution of the energy density.
$\rho_H$ and $\rho_L$ are the energy density of the heavier and lighter modes and $\rho_\mathrm{mix}$ is their sum.
As a comparison, we also consider the case with $N = 0$, where $a$ and $\phi$ decouple from each other.
By solving the dynamics of each field with the initial Bunch-Davies distribution, we obtain the energy densities of $a$ and $\phi$, $\rho_{a,\mathrm{no}\text{-}\mathrm{mixing}}$ and $\rho_{\phi,\mathrm{no}\text{-}\mathrm{mixing}}$.
After $V_\mathrm{QCD}$ arises, the heavier mode is approximately $a$.
Since $a$ grows due to the slow roll in the $\Phi$-direction before oscillations, $\rho_H$ is enhanced compared with $\rho_{a,\mathrm{no}\text{-}\mathrm{mixing}}$.
On the other hand, $\rho_L$ is almost the same as $\rho_{\phi,\mathrm{no}\text{-}\mathrm{mixing}}$ since the slow roll of $\Phi$ or the oscillation of $a$ has little effect on the time evolution of $\phi$.
As a result, the total energy density is hardly enhanced.

\begin{figure}[!p]
    \begin{center}  
        \includegraphics[width=90mm]{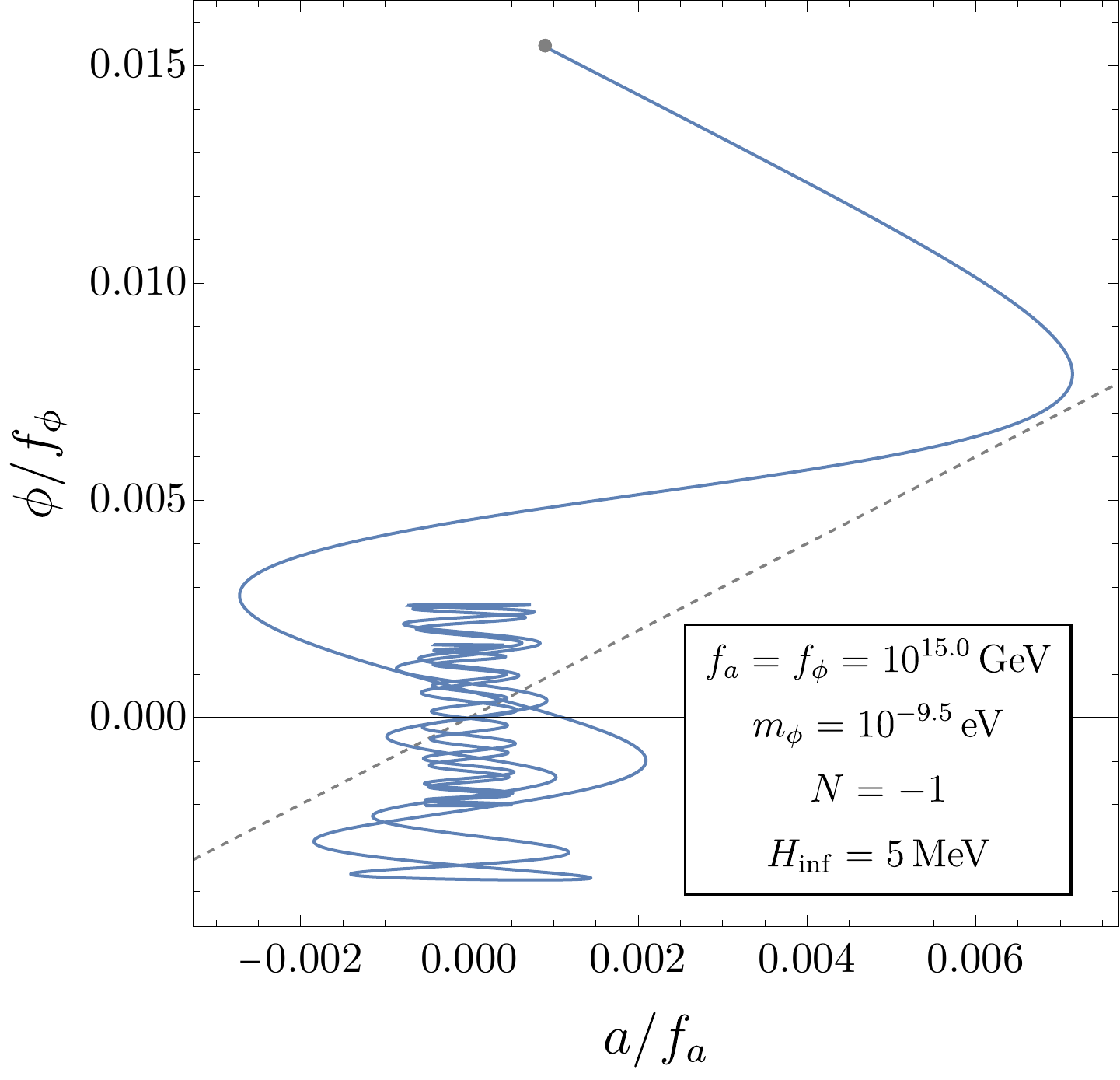}
        \\
        \vspace{5mm}
        \includegraphics[width=110mm]{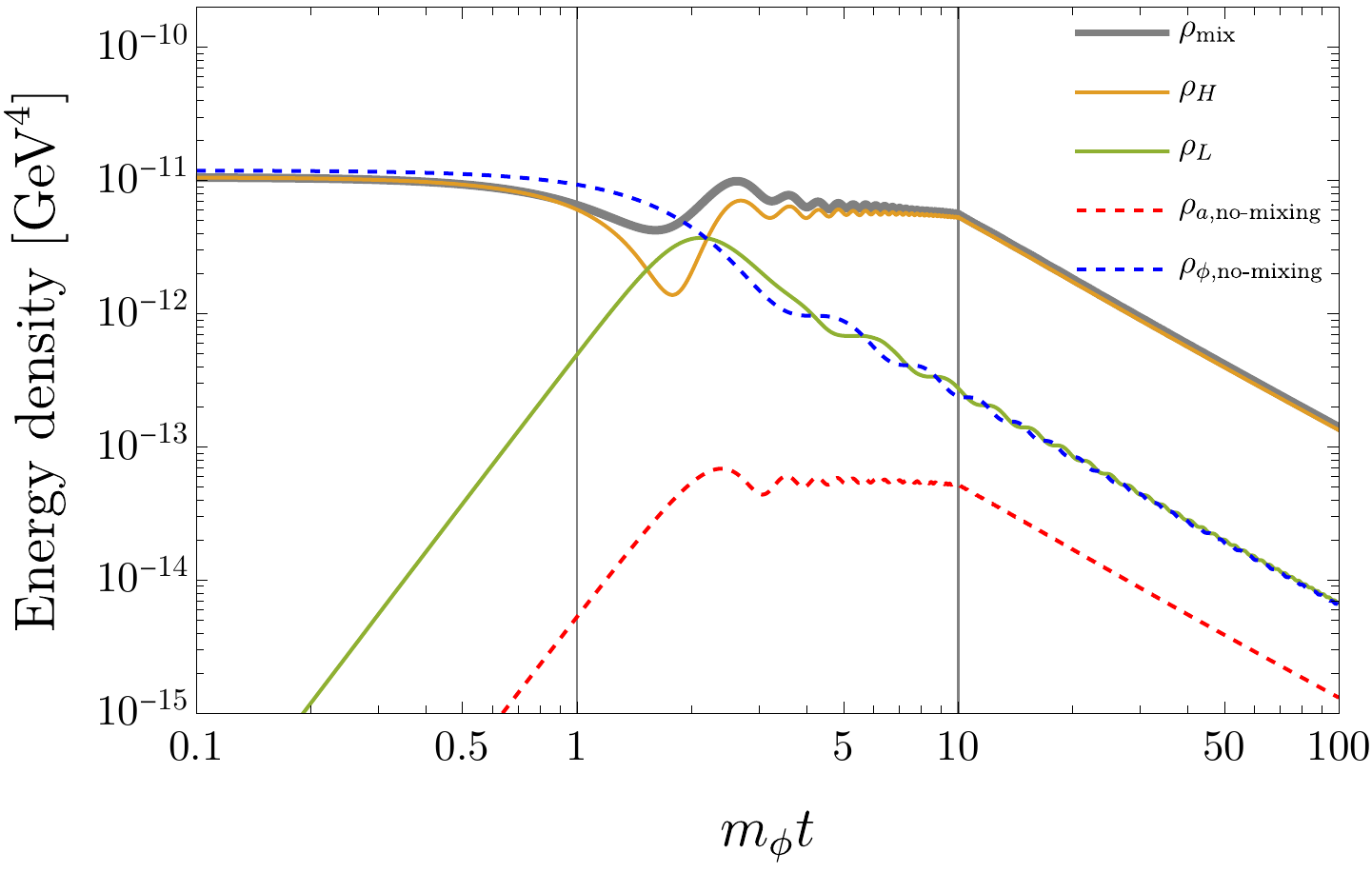}
        \end{center}
    \caption{%
        Same as Fig.~\ref{fig: evolution -10.5} but for $m_\phi = 10^{-9.5}$\,eV.
        The vertical lines in the bottom panel represent $m_\phi t = 1$ and $T = T_\mathrm{QCD}$ from left to right. 
        In contrast to the usual case with stochastic axions, the heavier axion (mostly the QCD axion) has a larger abundance due to the mixing effect.
    }
    \label{fig: evolution -9.5} 
\end{figure}

Next, we show the result for $m_\phi = 10^{-9.5}$\,eV in Fig.~\ref{fig: evolution -9.5}.
As before, $a/f_a$ is initially smaller than $\phi/f_\phi$ because of $m_{a0} > m_\phi$.
For $T \gg T_\mathrm{QCD}$, the potential is dominated by $V_{a \phi}$, and $\Phi$ starts to roll down the potential while $A$ remains constant.
As the temperature decreases, $V_\mathrm{QCD}$ becomes relevant and then dominant.
Thus, the field motion changes its direction, and $a$ starts to oscillate rapidly.
In this process, $a$ acquires a larger field value than the initial condition, and the energy density of the two fields is enhanced compared with the case where the two fields evolve independently.
In particular, $\rho_H$ is significantly enhanced compared with $\rho_{a,\mathrm{no}\text{-}\mathrm{mixing}}$ while $\rho_L$ is not so different from $\rho_{\phi,\mathrm{no}\text{-}\mathrm{mixing}}$.
As a result, the total energy density is also enhanced due to the interplay of the two fields.

\begin{figure}[!p]
    \begin{center}  
        \includegraphics[width=90mm]{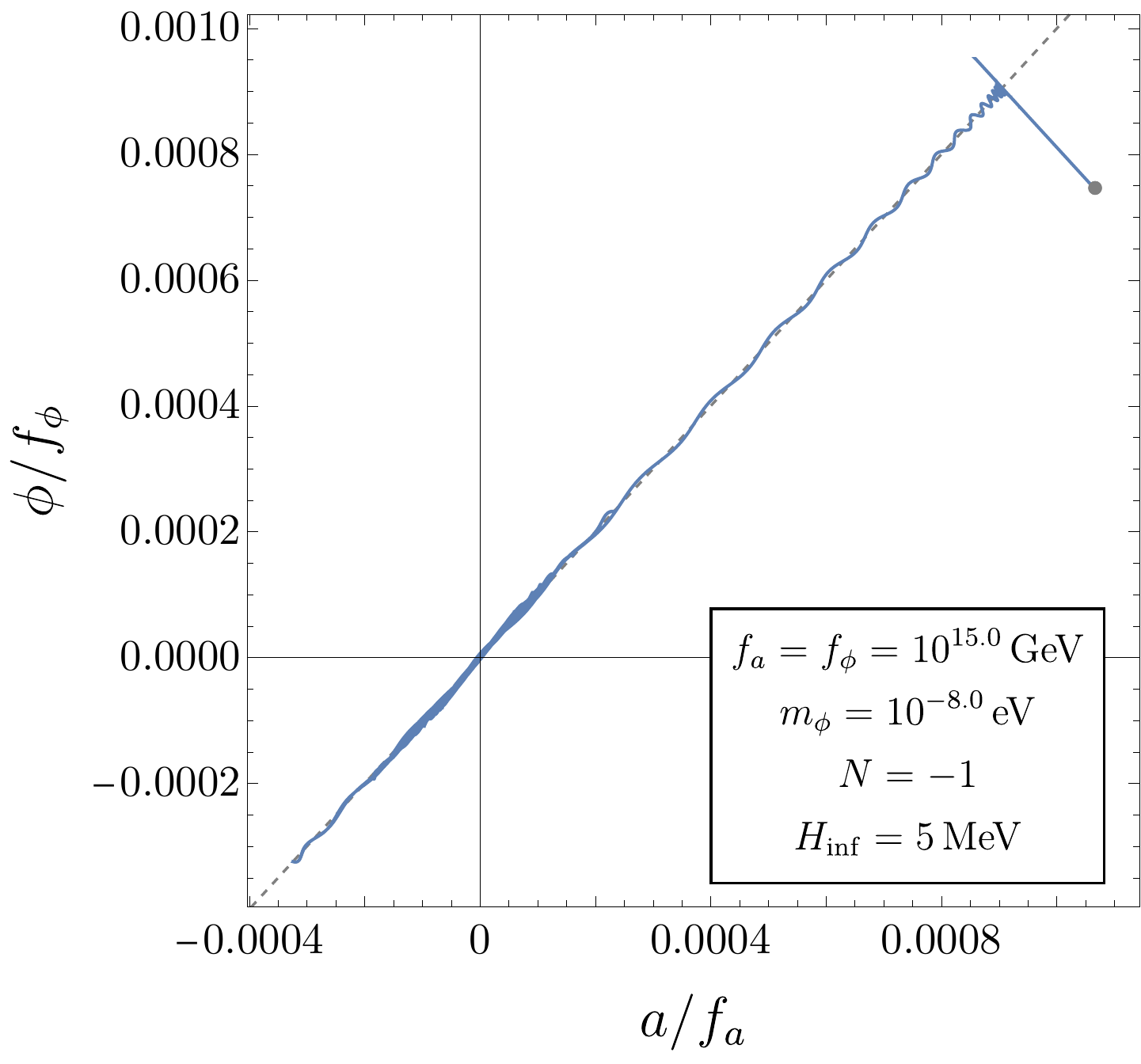}
        \\
        \vspace{5mm}
        \includegraphics[width=110mm]{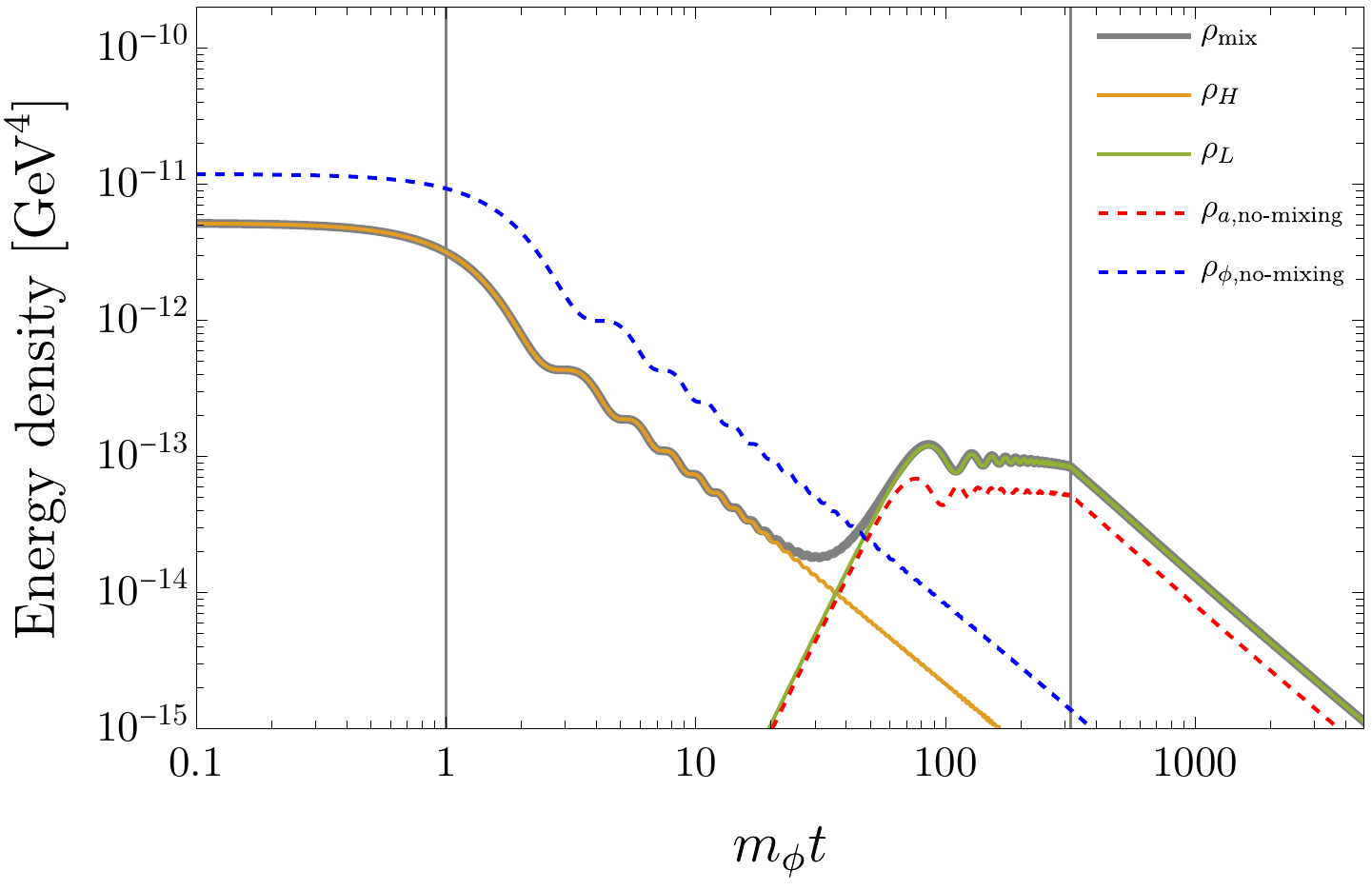}
        \end{center}
    \caption{%
        Same as Fig.~\ref{fig: evolution -10.5} but for $m_\phi = 10^{-8}$\,eV. The lighter axion (mostly the QCD axion) has a larger abundance. The slight enhancement over the unmixed case is due to the mixing effect on the decay constant, not the dynamics (see Fig.~\ref{fig: feff}).
    }
    \label{fig: evolution -8.0} 
\end{figure}

Finally, we show the result for $m_\phi = 10^{-8}$\,eV in Fig.~\ref{fig: evolution -8.0}.
In this case, the heavier mode $(\simeq \Phi)$ starts to oscillate and damps well before the emergence of $V_\mathrm{QCD}$.
Then, $\rho_L$ becomes dominant later.
As a result, the total energy density is different from the case with $N = 0$ only by a factor of $\simeq 1.6$.
This is because the lighter mode is approximately equal to $A$, which has an effective decay constant $f_{\mathrm{eff}, A} = \sqrt{2} f_a$ (see Eq.~\eqref{eq: effective decay constants}).
Considering the result of the standard misalignment mechanism for the QCD axion, $\Omega_a \propto f_a^{1.17}$~\cite{Bae:2008ue, Visinelli:2009zm,Ballesteros:2016xej}, we expect $\rho_L / \rho_{a,\mathrm{no}\text{-}\mathrm{mixing}} \simeq 2^{1.17/2} \simeq 1.5$.

\subsubsection{Enhancement in QCD axion abundance}

\begin{figure}[!t]
    \begin{center}  
        \includegraphics[width=105mm]{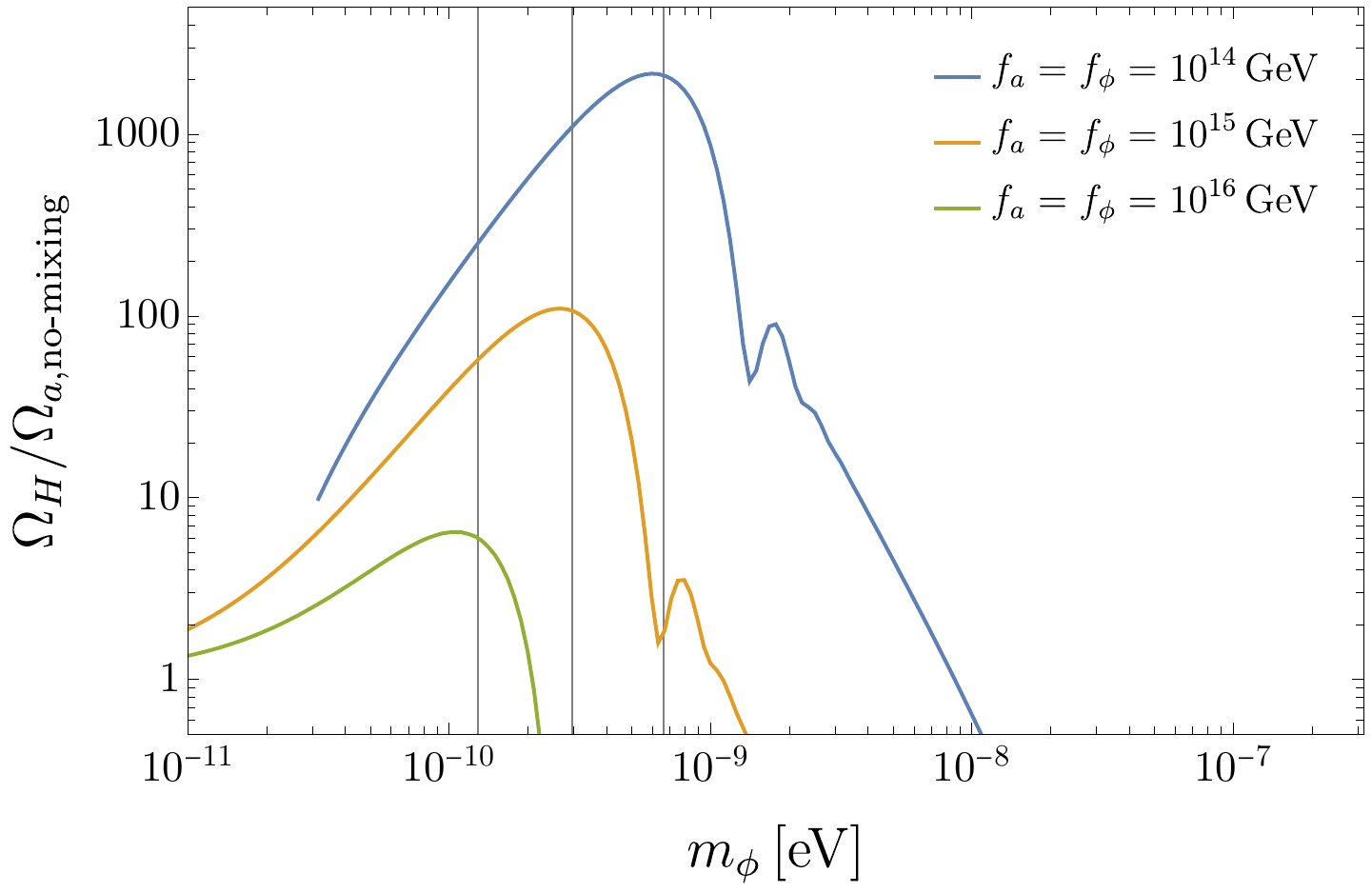}
        \end{center}
    \caption{%
        Ratio of $\Omega_H$ to $\Omega_{a,\mathrm{no}\text{-}\mathrm{mixing}}$.
        The vertical gray lines represent 
        $m_\phi = m_\mathrm{enh}$ for $f = 10^{14}$, $10^{15}$, and $10^{16}$\,GeV from right to left.
        Note that, in the region shown here, the heavier axion is identified with the QCD axion, which consists mainly of $a$.
    }
    \label{fig: enhancement H-a} 
\end{figure}

Next, we look at the $m_\phi$ dependence of the enhancement factor.
We show the ratio of the abundances of the heavier mode with mixing, $\Omega_H$, and the QCD axion without mixing, $\Omega_{a,\mathrm{no}\text{-}\mathrm{mixing}}$, for $f = 10^{14}$, $10^{15}$, and $10^{16}$\,GeV in Fig.~\ref{fig: enhancement H-a}.
Note that the heavier mode $s_H$ almost corresponds to the QCD axion $a$ at low temperatures for $m_{a0} \gg m_\phi$.
We see that $\Omega_H$ is enhanced around $m_\phi = m_\mathrm{enh}$ as expected.
On the other hand, the ratio becomes less than unity for larger $m_\phi$.
This is because $s_H$ starts to oscillate due to $V_{a \phi}$ earlier than $a$ without mixing for $m_\phi > m_\mathrm{enh}$.
The enhancement is most significant for $f = 10^{14}$\,GeV, with which the ratio is $\Omega_H/\Omega_{a,\mathrm{no}\text{-}\mathrm{mixing}}=\mathcal{O}(10^3)$ for $m_\phi \simeq m_\mathrm{enh}$.
The maximum ratio for $f = 10^{14}$\,GeV is larger than that for $f = 10^{16}$\,GeV by a factor of $333 \simeq 100^{1.26}$, which validates the relation obtained in Sec.~\ref{subsec: RD}, $\Omega_H / \Omega_{a,\mathrm{no}\text{-}\mathrm{mixing}} \propto f^{-1.34}$ as a rough estimate.

\begin{figure}[!t]
    \begin{center}  
        \includegraphics[width=105mm]{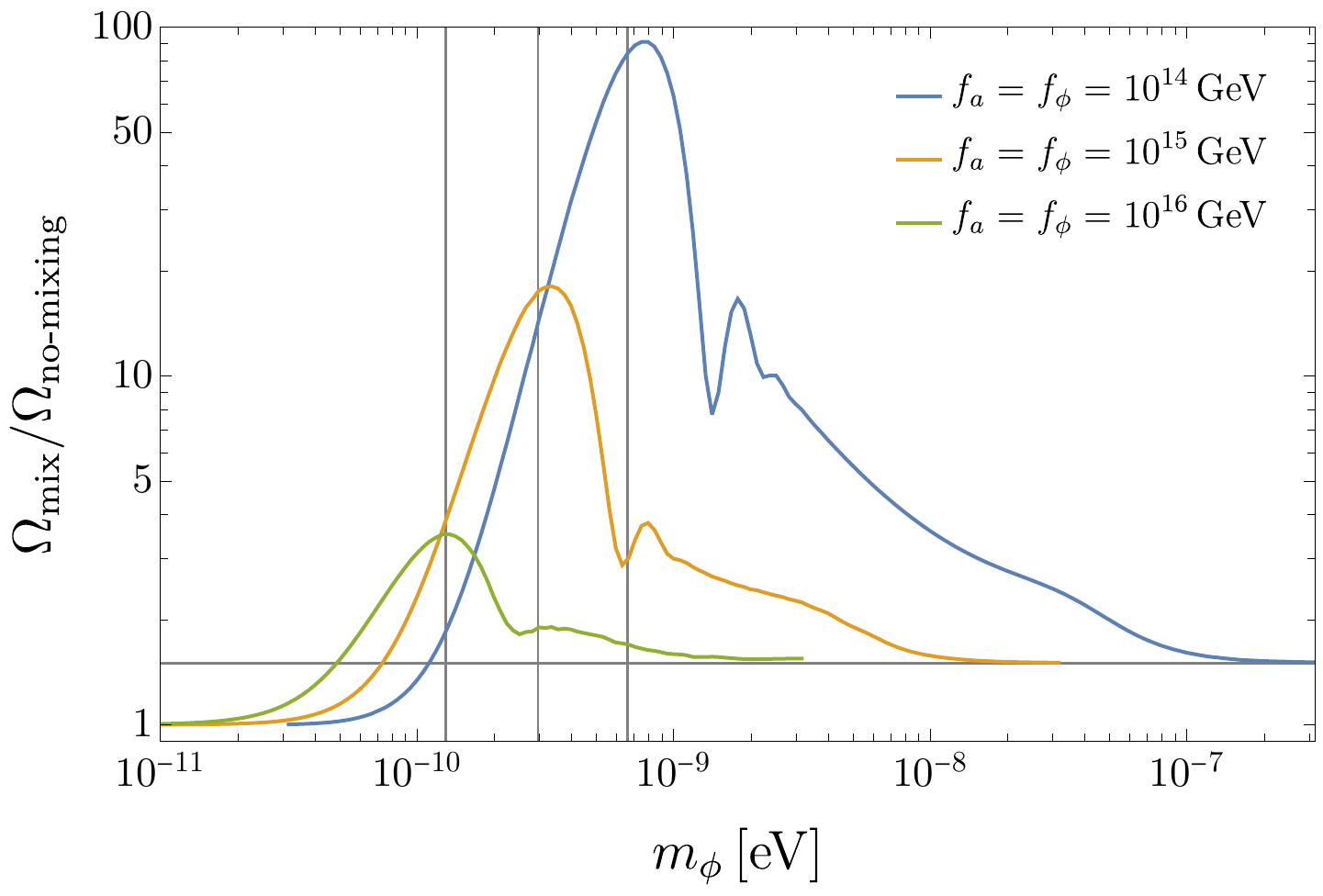}
        \end{center}
    \caption{%
        Enhancement of the energy density in the two-field dynamics compared with the sum of the one-field dynamics of $a$ and $\phi$.
        The vertical gray lines represent $m_\phi = m_\mathrm{enh}$ for $f = 10^{14}$, $10^{15}$, and $10^{16}$\,GeV from right to left.
        The horizontal gray line corresponds to $\Omega_\mathrm{mix}/\Omega_{\mathrm{no}\text{-}\mathrm{mixing}} = 2^{1.17/2} \simeq 1.5$.
    }
    \label{fig: enhancement} 
\end{figure}

We also show the ratio of the total abundances of the two fields between the cases with $N = -1$ and $N = 0$ against $m_\phi$ for $f = 10^{14}$, $10^{15}$, and $10^{16}$\,GeV in Fig.~\ref{fig: enhancement}.
The enhancement is most significant for $f = 10^{14}$\,GeV, with which the ratio is $\Omega_\mathrm{mix}/\Omega_{\mathrm{no}\text{-}\mathrm{mixing}}=\mathcal{O}(100)$ at the peak.
The wiggle in the right side of the peak corresponds to the oscillation phase of $\Phi$ when the QCD potential becomes relevant.
For smaller $m_\phi$, we obtain $\Omega_\mathrm{mix} \simeq \Omega_{\mathrm{no}\text{-}\mathrm{mixing}}$.
On the other hand, for larger $m_\phi$, we obtain $\Omega_\mathrm{mix} / \Omega_{\mathrm{no}\text{-}\mathrm{mixing}} \simeq 2^{1.17/2} \simeq 1.5$ as explained above.

\begin{figure}[!t]
    \begin{center}  
        \includegraphics[width=105mm]{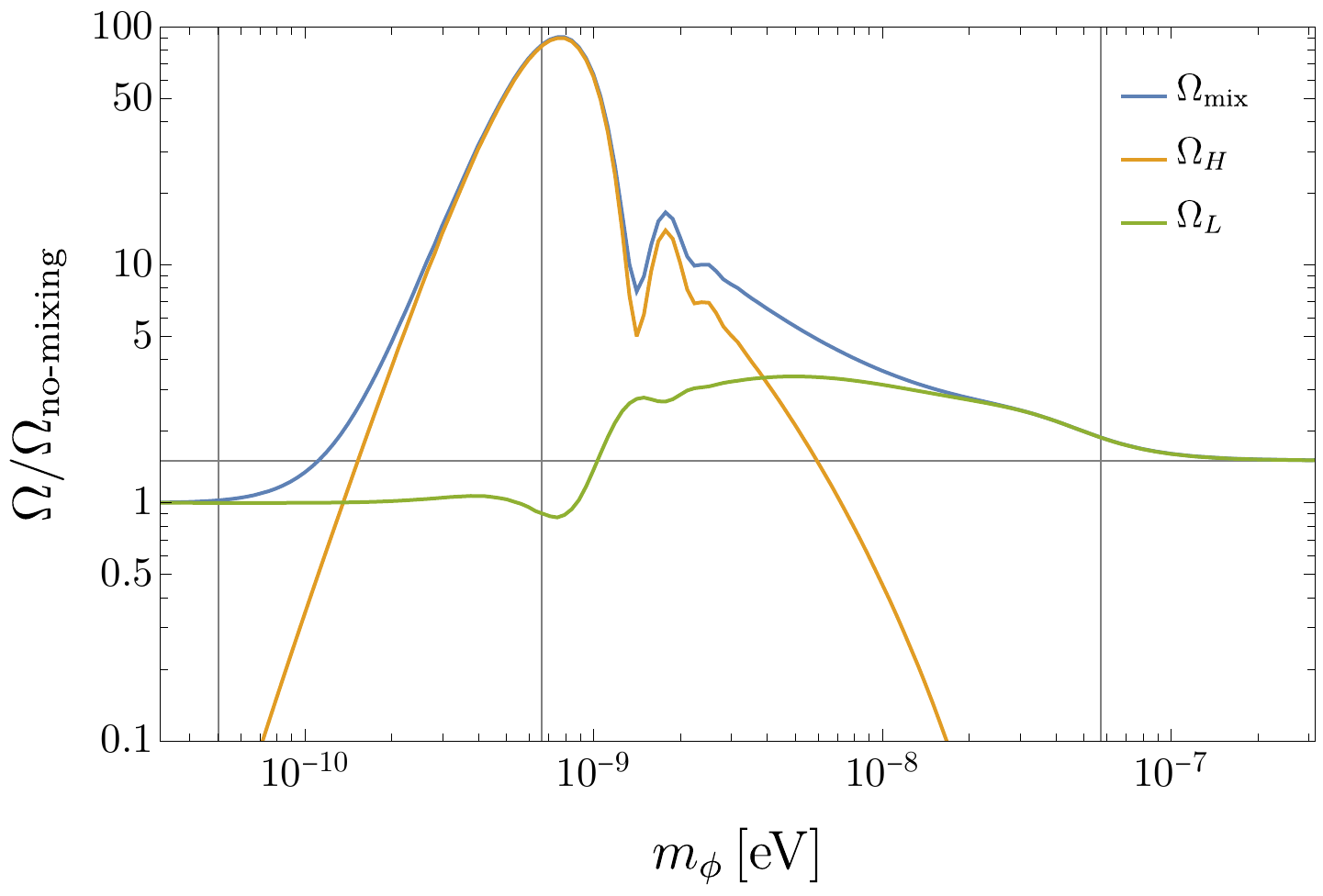}
        \end{center}
    \caption{%
        Enhancement of the energy density in the two-field dynamics compared with the sum of the single-field dynamics of $a$ and $\phi$ for $f = 10^{14}$\,GeV.
        The blue line is the same as in Fig.~\ref{fig: enhancement}.
        The orange and green lines represent the contributions of the heavier and lighter modes, respectively.
        The vertical gray lines represent $m_\phi = 3 H(T_\mathrm{QCD})$, $m_\mathrm{enh}$, and $m_{a0}$, from left to right.
    }
    \label{fig: H and L f14} 
\end{figure}

We have seen that $\rho_H$ becomes dominant when the enhancement is significant in Fig.~\ref{fig: evolution -9.5} in contrast to the case without enhancement in Figs.~\ref{fig: evolution -10.5} and \ref{fig: evolution -8.0}.
To visualize this trend, we show the contributions of the heavier and lighter modes to the enhancement in Fig.~\ref{fig: H and L f14}.
Here, we choose $f = 10^{14}$\,GeV, with which the enhancement is most significant in Figs.~\ref{fig: enhancement H-a} and \ref{fig: enhancement}.
We see that the heavier mode dominates the energy density for the mass region where the enhancement is significant.
In this mass region, the heavier mode corresponds to the QCD axion $a$ with $m_{a 0} \simeq 5.7 \times 10^{-8}$\,eV for $f = 10^{14}$\,GeV.
For $m_\phi \simeq 4.0 \times 10^{-9}$\,eV, both the heavier and lighter modes have the same order of energy densities larger than $\Omega_{\mathrm{no}\text{-}\mathrm{mixing}}$.
For $m_\phi \lesssim 3 H(T_\mathrm{QCD})$, the lighter mode $\simeq \phi$ is dominant, and, for $m_\phi \gtrsim m_{a 0}$, the lighter mode $\simeq A$ is dominant.

\subsubsection{Viable parameter space for dark matter}

If the axion potential can be approximated by mass terms, then the squares of the oscillation amplitudes are proportional to $H_\mathrm{inf}^4$ at the end of the inflation, and so are the energy densities at any epoch after inflation with the other parameters fixed.
Using this approximation, the energy density derived for $H_\mathrm{inf} = 5$\,MeV can be converted to $H_\mathrm{inf}$ such that the axion field explains all dark matter.
To see the validity of the mass approximation, we define the typical amplitude of the Bunch-Davies distribution in the $\phi$-direction:
\begin{align}
    \bar{\theta}_\phi
    \equiv 
    \sqrt{\frac{3}{8\pi^2}} \frac{H_\mathrm{inf}^2}{m_\phi f_\phi}
    \simeq
    0.19 \left( \frac{H_\mathrm{inf}}{10\,\mathrm{MeV}}\right)^2 
    \left( \frac{m_\phi}{10^{-9}\,\mathrm{eV}}\right)^{-1}
    \left( \frac{f_\phi}{10^{14}\,\mathrm{GeV}}\right)^{-1}
    \ .
\end{align}
Note that the typical misalignment angle in the $a$-direction is much smaller for the parameters of interest:
\begin{align}
    \bar{\theta}_a
    \equiv 
    \sqrt{\frac{3}{8\pi^2}} \frac{H_\mathrm{inf}^2}{m_{a 0} f_a}
    \simeq
    3.4 \times 10^{-3} \left( \frac{H_\mathrm{inf}}{10\,\mathrm{MeV}}\right)^2
    \ .
\end{align}
These typical angles, $\bar{\theta}_\phi$ and $\bar{\theta}_a$, correspond to the initial values of $\phi/f_\phi$ and $a/f_a$ with $c_H = c_L = 1$ in the limit of $m_{a 0} \gg m_\phi$.

\begin{figure}[!t]
    \begin{center}  
        \includegraphics[width=105mm]{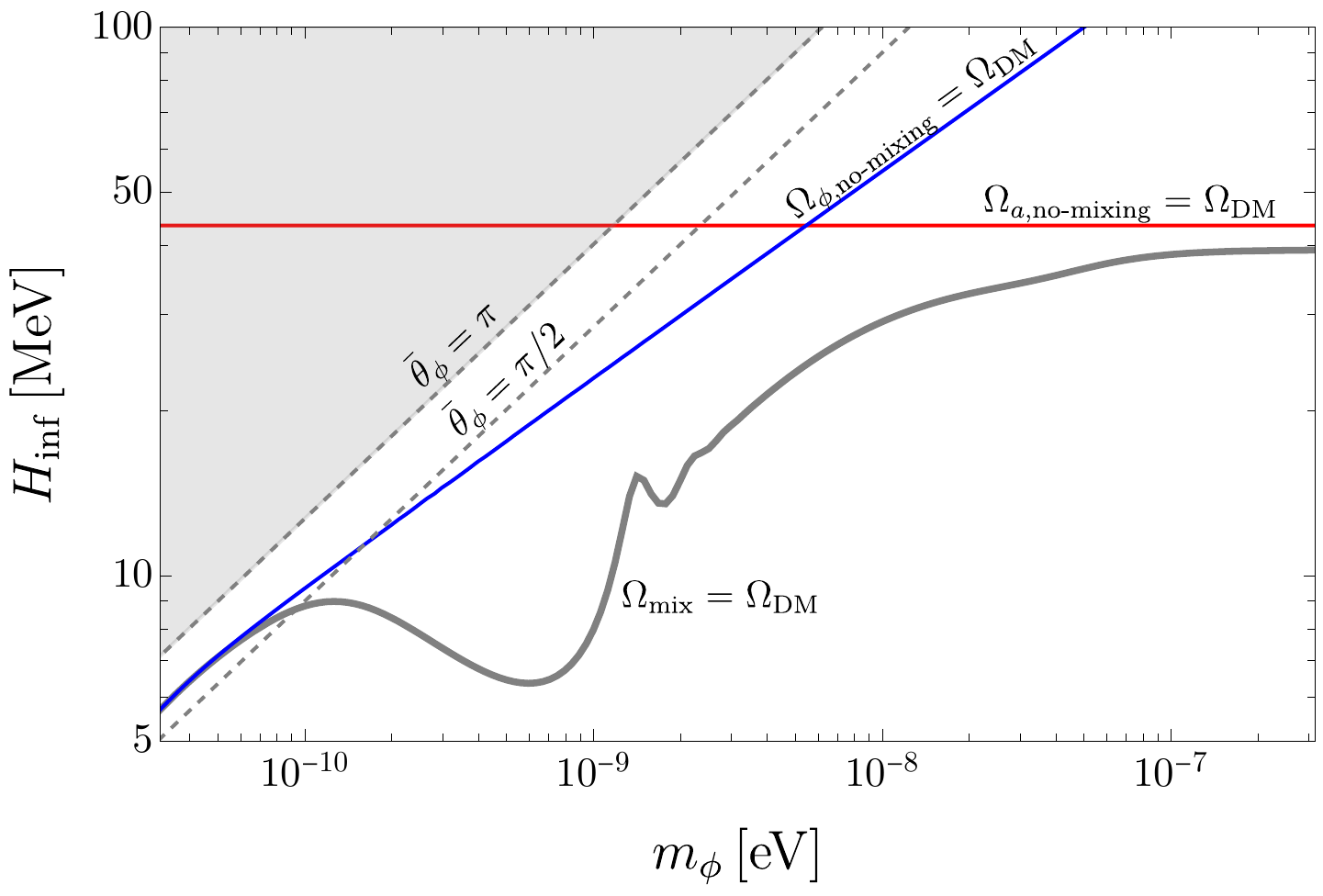}
        \end{center}
    \caption{%
        Hubble parameter during inflation with which the two fields with mixing (thick gray), $a$ without mixing (red), and $\phi$ without mixing (blue) explain all dark matter for $f = 10^{14}$\,GeV and $c_H = c_L = 1$.
        The gray dashed lines represent $\bar{\theta}_\phi = \pi/2$ and $\pi$. 
    }
    \label{fig: Hubble for DM f14} 
\end{figure}

We show $H_\mathrm{inf}$ explaining all dark matter for $f = 10^{14}$\,GeV and $c_H = c_L = 1$ in Fig.~\ref{fig: Hubble for DM f14}.
The solid lines represent $H_\mathrm{inf}$ with which the two fields with mixing (thick gray), $a$ without mixing (red), and $\phi$ without mixing (blue) explain all dark matter, respectively.
For $\bar{\theta}_\phi > \pi$ (the gray shaded region), the assumption that $a$ and $\phi$ oscillate around the origin $a = \phi = 0$ is invalid and our analysis cannot be applied directly.
In this region, however, the typical initial amplitude of $\phi$ is of order $f_\phi$, and the axion abundance is larger than the observed dark matter abundance.
For $\pi/2 < \bar{\theta}_\phi < \pi$ (between the two gray dashed lines), the mass approximation of the potential becomes inaccurate, and the result requires some correction.
We can see that the QCD axion can explain all dark matter with $H_{\rm inf}$ much smaller than the case without mixing effects.

\subsubsection{Initial condition dependence}

\begin{figure}[!t]
    \begin{center}  
        \includegraphics[width=105mm]{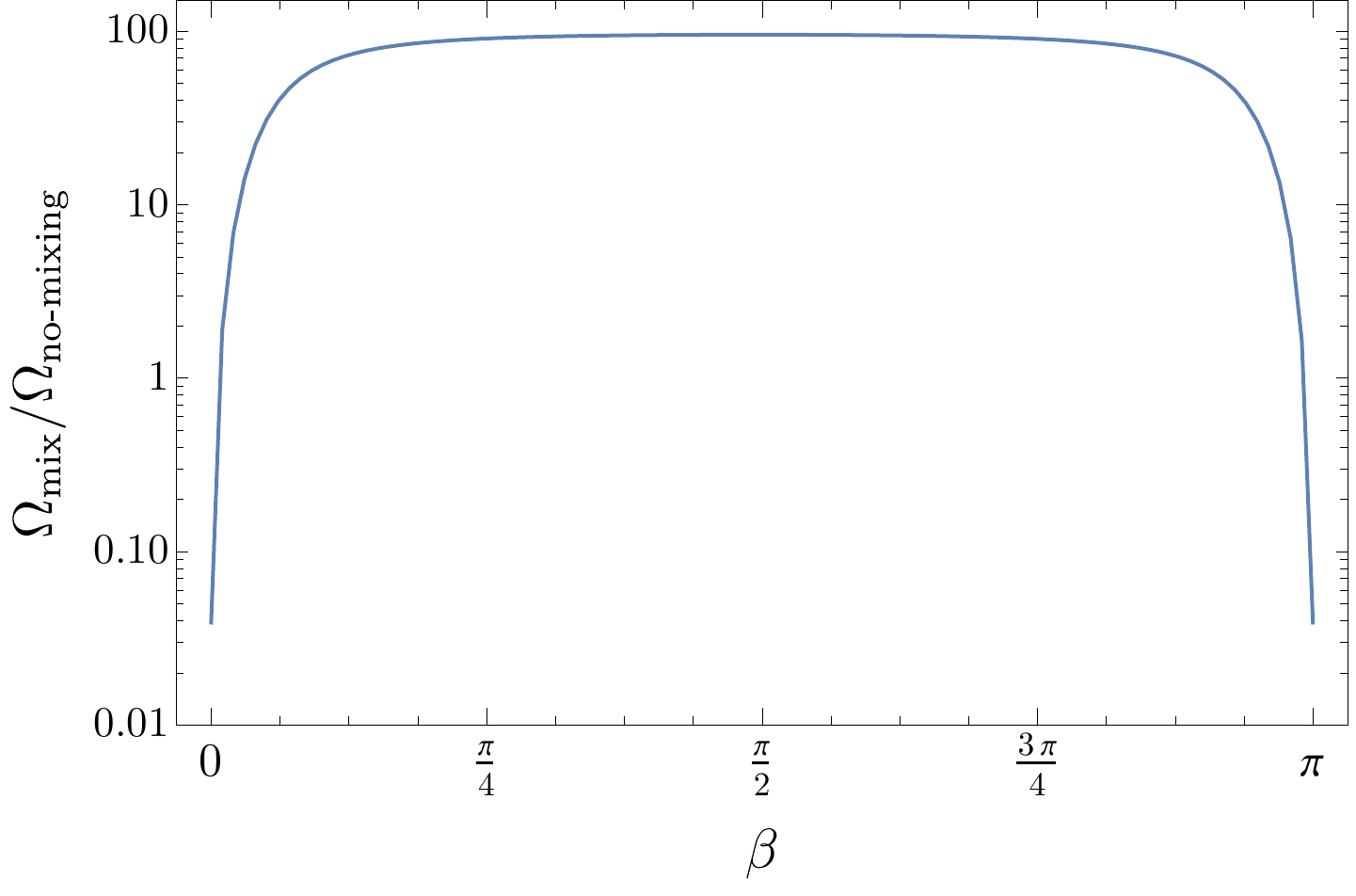}
        \end{center}
    \caption{%
        Dependence of the enhancement factor on the initial condition for $f = 10^{14}$\,GeV and $m_\phi = 7.9 \times 10^{-10}$\,eV.
    }
    \label{fig: enhance cHcL} 
\end{figure}

So far, we have taken $c_H = c_L = 1$.
Here, we discuss the importance of the initial conditions focusing on two parameter sets.
First, we consider the peak of enhancement:
\begin{align}
    f = 10^{14}\,\mathrm{GeV}
    \ , \quad 
    m_\phi = 7.9 \times 10^{-10}
    \,\mathrm{eV}
    \ ,
\end{align}
and investigate how the enhancement depends on the initial condition.
As long as the mass approximation of the potential is valid, the field dynamics is linear and the energy density is proportional to the square of the field amplitudes. 
Thus, the enhancement factor $\Omega_\mathrm{mix}/\Omega_{\mathrm{no}\text{-}\mathrm{mixing}}$ depends on the initial condition through $c_H/c_L$.
We parameterize the initial condition by $0 \leq \beta < \pi$ as
\begin{align}
    c_H = \cos \beta 
    \ , \quad 
    c_L = \sin \beta 
    \ ,
\end{align}
and perform the numerical simulations for both $N = -1$ and $N = 0$.
Note that $V(a, \phi)$ is invariant with $(a, \phi) \to (-a, -\phi)$ and that $\pi \leq \beta < 2\pi$ leads to the same energy densities as $0 \leq \beta < \pi$.
We show the dependence of the enhancement factor on $\beta$ in Fig.~\ref{fig: enhance cHcL}.
The enhancement is suppressed around $\beta = 0$ and $\pi$, where $c_L$ is small compared with $|c_H|$.
This behavior can be understood by the observation that the enhancement is due to the conversion of $\phi$ to $a$ in the early stage of field oscillations (see Fig.~\ref{fig: evolution -9.5}). 
For typical initial conditions, $\phi_\mathrm{init}/f_\phi$ is larger than $a_\mathrm{init}/f_a$, and the motion in the $\Phi$-direction enhances $a/f_a$.
On the other hand, for $c_L \ll |c_H|$, $\phi_\mathrm{init}/f_\phi \lesssim a_\mathrm{init}/f_a$ and the enhancement does not occur.

\begin{figure}[!t]
    \begin{center}  
        \includegraphics[width=105mm]{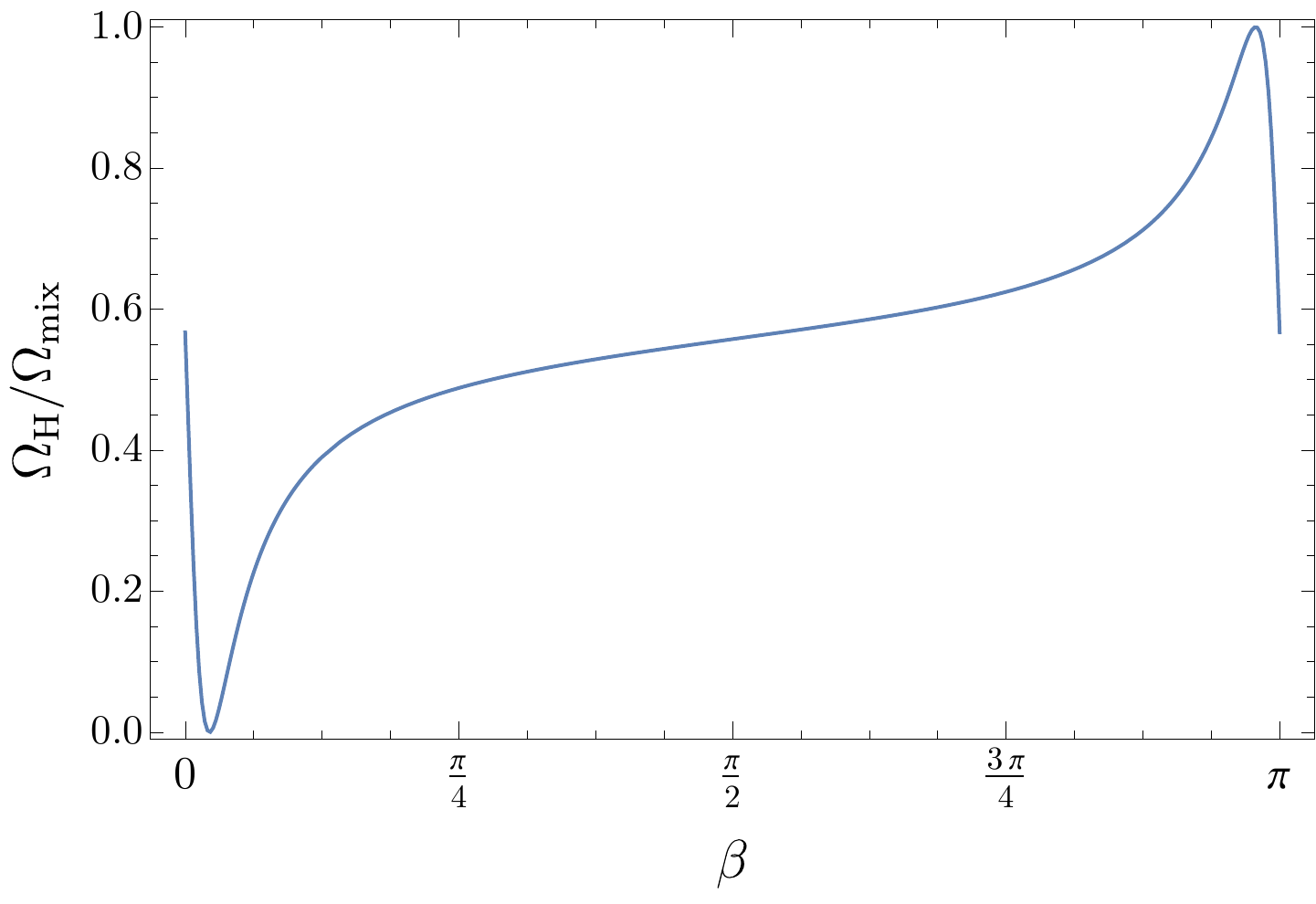}
        \end{center}
    \caption{%
        Dependence of the fraction of the energy density of the heavier mode on the initial condition for $f = 10^{14}$\,GeV and $m_\phi = 4.0 \times 10^{-9}$\,eV.
    }
    \label{fig: H ratio cHcL} 
\end{figure}

Next, we consider $m_\phi$ with which $\Omega_H \simeq \Omega_L$ in Fig.~\ref{fig: H and L f14}:
\begin{align}
    f = 10^{14}\,\mathrm{GeV}
    \ , \quad 
    m_\phi = 4.0 \times 10^{-9}
    \,\mathrm{eV}
    \ .
    \label{eq: m and f for H~L}
\end{align}
We show the energy ratio of the heavier mode, $\Omega_H / \Omega_\mathrm{mix}$, in Fig.~\ref{fig: H ratio cHcL}.
We see that both the heavier and lighter modes have non-negligible energy density except for two regions near $\beta = 0$ and $\pi$.
These two regions lead to $\Phi_\mathrm{init} \simeq 0$ and $A_\mathrm{init} \simeq 0$, respectively.
Then, one of $\rho_H$ and $\rho_L$ is highly suppressed as the initial condition because $(\Phi, A)$ correspond to $(s_H, s_L)$ at high temperatures $T \gg T_\mathrm{QCD}$.
Since the fields start oscillations before the emergence of $V_\mathrm{QCD}$ for the parameters in Eq.~\eqref{eq: m and f for H~L}, the energy densities of the heavier and lighter modes are not transferred to each other due to the resonant conversion~\cite{Kitajima:2014xla,Daido:2015cba,Ho:2018qur}.
As a result, one of the heavier and lighter modes becomes dominant for the two regions near $\beta = 0$ and $\pi$.

\section{Conclusions and discussions}
\label{sec: summary}

The stochastic axion scenario fits well with the string axiverse, where there are many axions with masses spread over a wide parameter range. 
As long as the inflationary scale is kept relatively small, all axions will stay near the potential minimum, thus avoiding the notorious cosmological moduli problem in the string axiverse.
In this paper, we have shown that among the axions in the axiverse, the QCD axion is special in the context of the stochastic axion scenario because it necessarily has a temperature-dependent potential.
Even if the axions have suppressed initial misalignment angles in the stochastic scenario, 
the mixing between the QCD axion and another axion and the time dependence of the QCD axion potential make the field trajectory after inflation quite non-trivial. Especially when the mixing is non-resonant, the two axion exhibits a highly complicated behavior. 
Through this dynamics, the QCD axion abundance can be enhanced by many orders of magnitude compared to the case where mixing is neglected. 

Let us see if the QCD axion can be a dominant component of dark matter in the axiverse with stochastic axions.
In this scenario, the lighter axion tends to have a greater abundance for the same decay constant.
This is because while the energy density at the onset of oscillations is of the order of $H_{\rm inf}^4$, the lighter axion starts oscillating later.
On the other hand, the initial amplitude cannot exceed the decay constant, and therefore there is a lower bound on both $H_{\rm inf}$ and the axion mass $m_\phi$ to account for all dark matter by the (lightest) axion in this scenario, as shown in Ref.~\cite{Ho:2019ayl}.
For example, for $f_\phi = 10^{16}(10^{14})$\,GeV, the lower bound is $H_{\rm inf} \gtrsim 10$\,keV ($10$\,MeV) and $m_\phi \gtrsim 10^{-18} (10^{-10})\,$eV.
Now, from Fig.~\ref{fig: Hubble for DM f14}, one can see that the QCD axion can explain all the dark matter for $H_{\rm inf} \simeq 7$ MeV and $m_\phi \simeq 10^{-9}$\,eV with mixing.
Thus, if the decay constant for other axions is universal and equal to $10^{14}$\,GeV, the contributions of the other axions are subdominant, i.e., the QCD axion is the dominant component of dark matter in the string axiverse scenario. 
This decay constant is somewhat lower than those conventionally adopted in the string axiverse, but could potentially be realized through a large volume compactification scenario~\cite{Balasubramanian:2005zx}.
We also note that the decay constant can be larger than  $10^{14}$\,GeV, if there is no axion near the lower bound on the mass.

Inflation with $H_{\rm inf}$ of order MeV is a low-scale inflation, but it is high enough for successful cosmology. This is because the corresponding energy scale of the inflaton potential is about $10^{7\text{--}8}$\,GeV, and the reheating temperature can be significantly higher than the weak scale. Consequently, we could use the active sphaleron reaction along with several potential baryogenesis mechanisms, such as leptogenesis and electroweak baryogenesis.

Interestingly, the QCD axion dark matter with the decay constant of ${\cal O}(10^{14})$\,GeV can be searched for through e.g., lumped element experiments~\cite{Silva-Feaver:2016qhh,Gramolin:2020ict,DMRadio:2022pkf,Berlin:2020vrk}. 
If the non-trivial dynamics was caused by the mixing between the QCD axion and another axion, there should be another axion in the mass range of $10^{-11}$\,eV to the mass of the QCD axion. The existence of such an axion with a mass close to that of the QCD axion has been discussed in various contexts~\cite{Kitajima:2014xla,Daido:2015cba,Ho:2018qur,Chen:2021hfq,Gavela:2023tzu}, and if we can find both of them, such an axiverse scenario with the stochastic axions would be one of the plausible possibilities.

\section*{Acknowledgments}
This work is supported by JSPS Core-to-Core Program (grant number: JPJSCCA20200002) (F.T.), JSPS KAKENHI Grant Numbers JP23KJ0088 (K.M.), 20H01894 (F.T.), 20H05851 (F.T. and W.Y.), 21K20364 (W.Y.), 22K14029 (W.Y.), and 22H01215 (W.Y.).
This article is based upon work from COST Action COSMIC WISPers CA21106, supported by COST (European Cooperation in Science and Technology).

\bibliographystyle{apsrev4-1}
\bibliography{ref}

\begin{thebibliography}{88}%
\makeatletter
\providecommand \@ifxundefined [1]{%
 \@ifx{#1\undefined}
}%
\providecommand \@ifnum [1]{%
 \ifnum #1\expandafter \@firstoftwo
 \else \expandafter \@secondoftwo
 \fi
}%
\providecommand \@ifx [1]{%
 \ifx #1\expandafter \@firstoftwo
 \else \expandafter \@secondoftwo
 \fi
}%
\providecommand \natexlab [1]{#1}%
\providecommand \enquote  [1]{``#1''}%
\providecommand \bibnamefont  [1]{#1}%
\providecommand \bibfnamefont [1]{#1}%
\providecommand \citenamefont [1]{#1}%
\providecommand \href@noop [0]{\@secondoftwo}%
\providecommand \href [0]{\begingroup \@sanitize@url \@href}%
\providecommand \@href[1]{\@@startlink{#1}\@@href}%
\providecommand \@@href[1]{\endgroup#1\@@endlink}%
\providecommand \@sanitize@url [0]{\catcode `\\12\catcode `\$12\catcode
  `\&12\catcode `\#12\catcode `\^12\catcode `\_12\catcode `\%12\relax}%
\providecommand \@@startlink[1]{}%
\providecommand \@@endlink[0]{}%
\providecommand \url  [0]{\begingroup\@sanitize@url \@url }%
\providecommand \@url [1]{\endgroup\@href {#1}{\urlprefix }}%
\providecommand \urlprefix  [0]{URL }%
\providecommand \Eprint [0]{\href }%
\providecommand \doibase [0]{http://dx.doi.org/}%
\providecommand \selectlanguage [0]{\@gobble}%
\providecommand \bibinfo  [0]{\@secondoftwo}%
\providecommand \bibfield  [0]{\@secondoftwo}%
\providecommand \translation [1]{[#1]}%
\providecommand \BibitemOpen [0]{}%
\providecommand \bibitemStop [0]{}%
\providecommand \bibitemNoStop [0]{.\EOS\space}%
\providecommand \EOS [0]{\spacefactor3000\relax}%
\providecommand \BibitemShut  [1]{\csname bibitem#1\endcsname}%
\let\auto@bib@innerbib\@empty
\bibitem [{\citenamefont {Witten}(1984)}]{Witten:1984dg}%
  \BibitemOpen
  \bibfield  {author} {\bibinfo {author} {\bibfnamefont {E.}~\bibnamefont
  {Witten}},\ }\href {\doibase 10.1016/0370-2693(84)90422-2} {\bibfield
  {journal} {\bibinfo  {journal} {Phys. Lett. B}\ }\textbf {\bibinfo {volume}
  {149}},\ \bibinfo {pages} {351} (\bibinfo {year} {1984})}\BibitemShut
  {NoStop}%
\bibitem [{\citenamefont {Svrcek}\ and\ \citenamefont
  {Witten}(2006)}]{Svrcek:2006yi}%
  \BibitemOpen
  \bibfield  {author} {\bibinfo {author} {\bibfnamefont {P.}~\bibnamefont
  {Svrcek}}\ and\ \bibinfo {author} {\bibfnamefont {E.}~\bibnamefont
  {Witten}},\ }\href {\doibase 10.1088/1126-6708/2006/06/051} {\bibfield
  {journal} {\bibinfo  {journal} {JHEP}\ }\textbf {\bibinfo {volume} {06}},\
  \bibinfo {pages} {051} (\bibinfo {year} {2006})},\ \Eprint
  {http://arxiv.org/abs/hep-th/0605206} {arXiv:hep-th/0605206} \BibitemShut
  {NoStop}%
\bibitem [{\citenamefont {Conlon}(2006)}]{Conlon:2006tq}%
  \BibitemOpen
  \bibfield  {author} {\bibinfo {author} {\bibfnamefont {J.~P.}\ \bibnamefont
  {Conlon}},\ }\href {\doibase 10.1088/1126-6708/2006/05/078} {\bibfield
  {journal} {\bibinfo  {journal} {JHEP}\ }\textbf {\bibinfo {volume} {05}},\
  \bibinfo {pages} {078} (\bibinfo {year} {2006})},\ \Eprint
  {http://arxiv.org/abs/hep-th/0602233} {arXiv:hep-th/0602233} \BibitemShut
  {NoStop}%
\bibitem [{\citenamefont {Arvanitaki}\ \emph {et~al.}(2010)\citenamefont
  {Arvanitaki}, \citenamefont {Dimopoulos}, \citenamefont {Dubovsky},
  \citenamefont {Kaloper},\ and\ \citenamefont
  {March-Russell}}]{Arvanitaki:2009fg}%
  \BibitemOpen
  \bibfield  {author} {\bibinfo {author} {\bibfnamefont {A.}~\bibnamefont
  {Arvanitaki}}, \bibinfo {author} {\bibfnamefont {S.}~\bibnamefont
  {Dimopoulos}}, \bibinfo {author} {\bibfnamefont {S.}~\bibnamefont
  {Dubovsky}}, \bibinfo {author} {\bibfnamefont {N.}~\bibnamefont {Kaloper}}, \
  and\ \bibinfo {author} {\bibfnamefont {J.}~\bibnamefont {March-Russell}},\
  }\href {\doibase 10.1103/PhysRevD.81.123530} {\bibfield  {journal} {\bibinfo
  {journal} {Phys. Rev. D}\ }\textbf {\bibinfo {volume} {81}},\ \bibinfo
  {pages} {123530} (\bibinfo {year} {2010})},\ \Eprint
  {http://arxiv.org/abs/0905.4720} {arXiv:0905.4720 [hep-th]} \BibitemShut
  {NoStop}%
\bibitem [{\citenamefont {Acharya}\ \emph {et~al.}(2010)\citenamefont
  {Acharya}, \citenamefont {Bobkov},\ and\ \citenamefont
  {Kumar}}]{Acharya:2010zx}%
  \BibitemOpen
  \bibfield  {author} {\bibinfo {author} {\bibfnamefont {B.~S.}\ \bibnamefont
  {Acharya}}, \bibinfo {author} {\bibfnamefont {K.}~\bibnamefont {Bobkov}}, \
  and\ \bibinfo {author} {\bibfnamefont {P.}~\bibnamefont {Kumar}},\ }\href
  {\doibase 10.1007/JHEP11(2010)105} {\bibfield  {journal} {\bibinfo  {journal}
  {JHEP}\ }\textbf {\bibinfo {volume} {11}},\ \bibinfo {pages} {105} (\bibinfo
  {year} {2010})},\ \Eprint {http://arxiv.org/abs/1004.5138} {arXiv:1004.5138
  [hep-th]} \BibitemShut {NoStop}%
\bibitem [{\citenamefont {Higaki}\ and\ \citenamefont
  {Kobayashi}(2011)}]{Higaki:2011me}%
  \BibitemOpen
  \bibfield  {author} {\bibinfo {author} {\bibfnamefont {T.}~\bibnamefont
  {Higaki}}\ and\ \bibinfo {author} {\bibfnamefont {T.}~\bibnamefont
  {Kobayashi}},\ }\href {\doibase 10.1103/PhysRevD.84.045021} {\bibfield
  {journal} {\bibinfo  {journal} {Phys. Rev. D}\ }\textbf {\bibinfo {volume}
  {84}},\ \bibinfo {pages} {045021} (\bibinfo {year} {2011})},\ \Eprint
  {http://arxiv.org/abs/1106.1293} {arXiv:1106.1293 [hep-th]} \BibitemShut
  {NoStop}%
\bibitem [{\citenamefont {Cicoli}\ \emph {et~al.}(2012)\citenamefont {Cicoli},
  \citenamefont {Goodsell},\ and\ \citenamefont {Ringwald}}]{Cicoli:2012sz}%
  \BibitemOpen
  \bibfield  {author} {\bibinfo {author} {\bibfnamefont {M.}~\bibnamefont
  {Cicoli}}, \bibinfo {author} {\bibfnamefont {M.}~\bibnamefont {Goodsell}}, \
  and\ \bibinfo {author} {\bibfnamefont {A.}~\bibnamefont {Ringwald}},\ }\href
  {\doibase 10.1007/JHEP10(2012)146} {\bibfield  {journal} {\bibinfo  {journal}
  {JHEP}\ }\textbf {\bibinfo {volume} {10}},\ \bibinfo {pages} {146} (\bibinfo
  {year} {2012})},\ \Eprint {http://arxiv.org/abs/1206.0819} {arXiv:1206.0819
  [hep-th]} \BibitemShut {NoStop}%
\bibitem [{\citenamefont {Demirtas}\ \emph {et~al.}(2020)\citenamefont
  {Demirtas}, \citenamefont {Long}, \citenamefont {McAllister},\ and\
  \citenamefont {Stillman}}]{Demirtas:2018akl}%
  \BibitemOpen
  \bibfield  {author} {\bibinfo {author} {\bibfnamefont {M.}~\bibnamefont
  {Demirtas}}, \bibinfo {author} {\bibfnamefont {C.}~\bibnamefont {Long}},
  \bibinfo {author} {\bibfnamefont {L.}~\bibnamefont {McAllister}}, \ and\
  \bibinfo {author} {\bibfnamefont {M.}~\bibnamefont {Stillman}},\ }\href
  {\doibase 10.1007/JHEP04(2020)138} {\bibfield  {journal} {\bibinfo  {journal}
  {JHEP}\ }\textbf {\bibinfo {volume} {04}},\ \bibinfo {pages} {138} (\bibinfo
  {year} {2020})},\ \Eprint {http://arxiv.org/abs/1808.01282} {arXiv:1808.01282
  [hep-th]} \BibitemShut {NoStop}%
\bibitem [{\citenamefont {Marsh}\ and\ \citenamefont
  {Yin}(2021)}]{Marsh:2019bjr}%
  \BibitemOpen
  \bibfield  {author} {\bibinfo {author} {\bibfnamefont {D.~J.~E.}\
  \bibnamefont {Marsh}}\ and\ \bibinfo {author} {\bibfnamefont
  {W.}~\bibnamefont {Yin}},\ }\href {\doibase 10.1007/JHEP01(2021)169}
  {\bibfield  {journal} {\bibinfo  {journal} {JHEP}\ }\textbf {\bibinfo
  {volume} {01}},\ \bibinfo {pages} {169} (\bibinfo {year} {2021})},\ \Eprint
  {http://arxiv.org/abs/1912.08188} {arXiv:1912.08188 [hep-ph]} \BibitemShut
  {NoStop}%
\bibitem [{\citenamefont {Mehta}\ \emph {et~al.}(2020)\citenamefont {Mehta},
  \citenamefont {Demirtas}, \citenamefont {Long}, \citenamefont {Marsh},
  \citenamefont {Mcallister},\ and\ \citenamefont {Stott}}]{Mehta:2020kwu}%
  \BibitemOpen
  \bibfield  {author} {\bibinfo {author} {\bibfnamefont {V.~M.}\ \bibnamefont
  {Mehta}}, \bibinfo {author} {\bibfnamefont {M.}~\bibnamefont {Demirtas}},
  \bibinfo {author} {\bibfnamefont {C.}~\bibnamefont {Long}}, \bibinfo {author}
  {\bibfnamefont {D.~J.~E.}\ \bibnamefont {Marsh}}, \bibinfo {author}
  {\bibfnamefont {L.}~\bibnamefont {Mcallister}}, \ and\ \bibinfo {author}
  {\bibfnamefont {M.~J.}\ \bibnamefont {Stott}},\ }\href@noop {} {\  (\bibinfo
  {year} {2020})},\ \Eprint {http://arxiv.org/abs/2011.08693} {arXiv:2011.08693
  [hep-th]} \BibitemShut {NoStop}%
\bibitem [{\citenamefont {Chen}\ \emph {et~al.}(2022)\citenamefont {Chen},
  \citenamefont {Kobakhidze}, \citenamefont {O'Hare}, \citenamefont {Picker},\
  and\ \citenamefont {Pierobon}}]{Chen:2021hfq}%
  \BibitemOpen
  \bibfield  {author} {\bibinfo {author} {\bibfnamefont {Z.}~\bibnamefont
  {Chen}}, \bibinfo {author} {\bibfnamefont {A.}~\bibnamefont {Kobakhidze}},
  \bibinfo {author} {\bibfnamefont {C.~A.~J.}\ \bibnamefont {O'Hare}}, \bibinfo
  {author} {\bibfnamefont {Z.~S.~C.}\ \bibnamefont {Picker}}, \ and\ \bibinfo
  {author} {\bibfnamefont {G.}~\bibnamefont {Pierobon}},\ }\href {\doibase
  10.1140/epjc/s10052-022-10909-6} {\bibfield  {journal} {\bibinfo  {journal}
  {Eur. Phys. J. C}\ }\textbf {\bibinfo {volume} {82}},\ \bibinfo {pages} {940}
  (\bibinfo {year} {2022})},\ \Eprint {http://arxiv.org/abs/2109.12920}
  {arXiv:2109.12920 [hep-ph]} \BibitemShut {NoStop}%
\bibitem [{\citenamefont {Mehta}\ \emph {et~al.}(2021)\citenamefont {Mehta},
  \citenamefont {Demirtas}, \citenamefont {Long}, \citenamefont {Marsh},
  \citenamefont {McAllister},\ and\ \citenamefont {Stott}}]{Mehta:2021pwf}%
  \BibitemOpen
  \bibfield  {author} {\bibinfo {author} {\bibfnamefont {V.~M.}\ \bibnamefont
  {Mehta}}, \bibinfo {author} {\bibfnamefont {M.}~\bibnamefont {Demirtas}},
  \bibinfo {author} {\bibfnamefont {C.}~\bibnamefont {Long}}, \bibinfo {author}
  {\bibfnamefont {D.~J.~E.}\ \bibnamefont {Marsh}}, \bibinfo {author}
  {\bibfnamefont {L.}~\bibnamefont {McAllister}}, \ and\ \bibinfo {author}
  {\bibfnamefont {M.~J.}\ \bibnamefont {Stott}},\ }\href {\doibase
  10.1088/1475-7516/2021/07/033} {\bibfield  {journal} {\bibinfo  {journal}
  {JCAP}\ }\textbf {\bibinfo {volume} {07}},\ \bibinfo {pages} {033} (\bibinfo
  {year} {2021})},\ \Eprint {http://arxiv.org/abs/2103.06812} {arXiv:2103.06812
  [hep-th]} \BibitemShut {NoStop}%
\bibitem [{\citenamefont {Cicoli}\ \emph {et~al.}(2022)\citenamefont {Cicoli},
  \citenamefont {Hebecker}, \citenamefont {Jaeckel},\ and\ \citenamefont
  {Wittner}}]{Cicoli:2022fzy}%
  \BibitemOpen
  \bibfield  {author} {\bibinfo {author} {\bibfnamefont {M.}~\bibnamefont
  {Cicoli}}, \bibinfo {author} {\bibfnamefont {A.}~\bibnamefont {Hebecker}},
  \bibinfo {author} {\bibfnamefont {J.}~\bibnamefont {Jaeckel}}, \ and\
  \bibinfo {author} {\bibfnamefont {M.}~\bibnamefont {Wittner}},\ }\href
  {\doibase 10.1007/JHEP09(2022)198} {\bibfield  {journal} {\bibinfo  {journal}
  {JHEP}\ }\textbf {\bibinfo {volume} {09}},\ \bibinfo {pages} {198} (\bibinfo
  {year} {2022})},\ \Eprint {http://arxiv.org/abs/2203.08833} {arXiv:2203.08833
  [hep-th]} \BibitemShut {NoStop}%
\bibitem [{\citenamefont {Peccei}\ and\ \citenamefont
  {Quinn}(1977{\natexlab{a}})}]{Peccei:1977hh}%
  \BibitemOpen
  \bibfield  {author} {\bibinfo {author} {\bibfnamefont {R.~D.}\ \bibnamefont
  {Peccei}}\ and\ \bibinfo {author} {\bibfnamefont {H.~R.}\ \bibnamefont
  {Quinn}},\ }\href {\doibase 10.1103/PhysRevLett.38.1440} {\bibfield
  {journal} {\bibinfo  {journal} {Phys. Rev. Lett.}\ }\textbf {\bibinfo
  {volume} {38}},\ \bibinfo {pages} {1440} (\bibinfo {year}
  {1977}{\natexlab{a}})}\BibitemShut {NoStop}%
\bibitem [{\citenamefont {Peccei}\ and\ \citenamefont
  {Quinn}(1977{\natexlab{b}})}]{Peccei:1977ur}%
  \BibitemOpen
  \bibfield  {author} {\bibinfo {author} {\bibfnamefont {R.~D.}\ \bibnamefont
  {Peccei}}\ and\ \bibinfo {author} {\bibfnamefont {H.~R.}\ \bibnamefont
  {Quinn}},\ }\href {\doibase 10.1103/PhysRevD.16.1791} {\bibfield  {journal}
  {\bibinfo  {journal} {Phys. Rev. D}\ }\textbf {\bibinfo {volume} {16}},\
  \bibinfo {pages} {1791} (\bibinfo {year} {1977}{\natexlab{b}})}\BibitemShut
  {NoStop}%
\bibitem [{\citenamefont {Weinberg}(1978)}]{Weinberg:1977ma}%
  \BibitemOpen
  \bibfield  {author} {\bibinfo {author} {\bibfnamefont {S.}~\bibnamefont
  {Weinberg}},\ }\href {\doibase 10.1103/PhysRevLett.40.223} {\bibfield
  {journal} {\bibinfo  {journal} {Phys. Rev. Lett.}\ }\textbf {\bibinfo
  {volume} {40}},\ \bibinfo {pages} {223} (\bibinfo {year} {1978})}\BibitemShut
  {NoStop}%
\bibitem [{\citenamefont {Wilczek}(1978)}]{Wilczek:1977pj}%
  \BibitemOpen
  \bibfield  {author} {\bibinfo {author} {\bibfnamefont {F.}~\bibnamefont
  {Wilczek}},\ }\href {\doibase 10.1103/PhysRevLett.40.279} {\bibfield
  {journal} {\bibinfo  {journal} {Phys. Rev. Lett.}\ }\textbf {\bibinfo
  {volume} {40}},\ \bibinfo {pages} {279} (\bibinfo {year} {1978})}\BibitemShut
  {NoStop}%
\bibitem [{\citenamefont {Carroll}(1998)}]{Carroll:1998zi}%
  \BibitemOpen
  \bibfield  {author} {\bibinfo {author} {\bibfnamefont {S.~M.}\ \bibnamefont
  {Carroll}},\ }\href {\doibase 10.1103/PhysRevLett.81.3067} {\bibfield
  {journal} {\bibinfo  {journal} {Phys. Rev. Lett.}\ }\textbf {\bibinfo
  {volume} {81}},\ \bibinfo {pages} {3067} (\bibinfo {year} {1998})},\ \Eprint
  {http://arxiv.org/abs/astro-ph/9806099} {arXiv:astro-ph/9806099} \BibitemShut
  {NoStop}%
\bibitem [{\citenamefont {Finelli}\ and\ \citenamefont
  {Galaverni}(2009)}]{Finelli:2008jv}%
  \BibitemOpen
  \bibfield  {author} {\bibinfo {author} {\bibfnamefont {F.}~\bibnamefont
  {Finelli}}\ and\ \bibinfo {author} {\bibfnamefont {M.}~\bibnamefont
  {Galaverni}},\ }\href {\doibase 10.1103/PhysRevD.79.063002} {\bibfield
  {journal} {\bibinfo  {journal} {Phys. Rev. D}\ }\textbf {\bibinfo {volume}
  {79}},\ \bibinfo {pages} {063002} (\bibinfo {year} {2009})},\ \Eprint
  {http://arxiv.org/abs/0802.4210} {arXiv:0802.4210 [astro-ph]} \BibitemShut
  {NoStop}%
\bibitem [{\citenamefont {Panda}\ \emph {et~al.}(2011)\citenamefont {Panda},
  \citenamefont {Sumitomo},\ and\ \citenamefont {Trivedi}}]{Panda:2010uq}%
  \BibitemOpen
  \bibfield  {author} {\bibinfo {author} {\bibfnamefont {S.}~\bibnamefont
  {Panda}}, \bibinfo {author} {\bibfnamefont {Y.}~\bibnamefont {Sumitomo}}, \
  and\ \bibinfo {author} {\bibfnamefont {S.~P.}\ \bibnamefont {Trivedi}},\
  }\href {\doibase 10.1103/PhysRevD.83.083506} {\bibfield  {journal} {\bibinfo
  {journal} {Phys. Rev. D}\ }\textbf {\bibinfo {volume} {83}},\ \bibinfo
  {pages} {083506} (\bibinfo {year} {2011})},\ \Eprint
  {http://arxiv.org/abs/1011.5877} {arXiv:1011.5877 [hep-th]} \BibitemShut
  {NoStop}%
\bibitem [{\citenamefont {Fedderke}\ \emph {et~al.}(2019)\citenamefont
  {Fedderke}, \citenamefont {Graham},\ and\ \citenamefont
  {Rajendran}}]{Fedderke:2019ajk}%
  \BibitemOpen
  \bibfield  {author} {\bibinfo {author} {\bibfnamefont {M.~A.}\ \bibnamefont
  {Fedderke}}, \bibinfo {author} {\bibfnamefont {P.~W.}\ \bibnamefont
  {Graham}}, \ and\ \bibinfo {author} {\bibfnamefont {S.}~\bibnamefont
  {Rajendran}},\ }\href {\doibase 10.1103/PhysRevD.100.015040} {\bibfield
  {journal} {\bibinfo  {journal} {Phys. Rev. D}\ }\textbf {\bibinfo {volume}
  {100}},\ \bibinfo {pages} {015040} (\bibinfo {year} {2019})},\ \Eprint
  {http://arxiv.org/abs/1903.02666} {arXiv:1903.02666 [astro-ph.CO]}
  \BibitemShut {NoStop}%
\bibitem [{\citenamefont {Fujita}\ \emph
  {et~al.}(2021{\natexlab{a}})\citenamefont {Fujita}, \citenamefont {Minami},
  \citenamefont {Murai},\ and\ \citenamefont {Nakatsuka}}]{Fujita:2020aqt}%
  \BibitemOpen
  \bibfield  {author} {\bibinfo {author} {\bibfnamefont {T.}~\bibnamefont
  {Fujita}}, \bibinfo {author} {\bibfnamefont {Y.}~\bibnamefont {Minami}},
  \bibinfo {author} {\bibfnamefont {K.}~\bibnamefont {Murai}}, \ and\ \bibinfo
  {author} {\bibfnamefont {H.}~\bibnamefont {Nakatsuka}},\ }\href {\doibase
  10.1103/PhysRevD.103.063508} {\bibfield  {journal} {\bibinfo  {journal}
  {Phys. Rev. D}\ }\textbf {\bibinfo {volume} {103}},\ \bibinfo {pages}
  {063508} (\bibinfo {year} {2021}{\natexlab{a}})},\ \Eprint
  {http://arxiv.org/abs/2008.02473} {arXiv:2008.02473 [astro-ph.CO]}
  \BibitemShut {NoStop}%
\bibitem [{\citenamefont {Fujita}\ \emph
  {et~al.}(2021{\natexlab{b}})\citenamefont {Fujita}, \citenamefont {Murai},
  \citenamefont {Nakatsuka},\ and\ \citenamefont {Tsujikawa}}]{Fujita:2020ecn}%
  \BibitemOpen
  \bibfield  {author} {\bibinfo {author} {\bibfnamefont {T.}~\bibnamefont
  {Fujita}}, \bibinfo {author} {\bibfnamefont {K.}~\bibnamefont {Murai}},
  \bibinfo {author} {\bibfnamefont {H.}~\bibnamefont {Nakatsuka}}, \ and\
  \bibinfo {author} {\bibfnamefont {S.}~\bibnamefont {Tsujikawa}},\ }\href
  {\doibase 10.1103/PhysRevD.103.043509} {\bibfield  {journal} {\bibinfo
  {journal} {Phys. Rev. D}\ }\textbf {\bibinfo {volume} {103}},\ \bibinfo
  {pages} {043509} (\bibinfo {year} {2021}{\natexlab{b}})},\ \Eprint
  {http://arxiv.org/abs/2011.11894} {arXiv:2011.11894 [astro-ph.CO]}
  \BibitemShut {NoStop}%
\bibitem [{\citenamefont {Takahashi}\ and\ \citenamefont
  {Yin}(2021)}]{Takahashi:2020tqv}%
  \BibitemOpen
  \bibfield  {author} {\bibinfo {author} {\bibfnamefont {F.}~\bibnamefont
  {Takahashi}}\ and\ \bibinfo {author} {\bibfnamefont {W.}~\bibnamefont
  {Yin}},\ }\href {\doibase 10.1088/1475-7516/2021/04/007} {\bibfield
  {journal} {\bibinfo  {journal} {JCAP}\ }\textbf {\bibinfo {volume} {04}},\
  \bibinfo {pages} {007} (\bibinfo {year} {2021})},\ \Eprint
  {http://arxiv.org/abs/2012.11576} {arXiv:2012.11576 [hep-ph]} \BibitemShut
  {NoStop}%
\bibitem [{\citenamefont {Nakagawa}\ \emph {et~al.}(2021)\citenamefont
  {Nakagawa}, \citenamefont {Takahashi},\ and\ \citenamefont
  {Yamada}}]{Nakagawa:2021nme}%
  \BibitemOpen
  \bibfield  {author} {\bibinfo {author} {\bibfnamefont {S.}~\bibnamefont
  {Nakagawa}}, \bibinfo {author} {\bibfnamefont {F.}~\bibnamefont {Takahashi}},
  \ and\ \bibinfo {author} {\bibfnamefont {M.}~\bibnamefont {Yamada}},\ }\href
  {\doibase 10.1103/PhysRevLett.127.181103} {\bibfield  {journal} {\bibinfo
  {journal} {Phys. Rev. Lett.}\ }\textbf {\bibinfo {volume} {127}},\ \bibinfo
  {pages} {181103} (\bibinfo {year} {2021})},\ \Eprint
  {http://arxiv.org/abs/2103.08153} {arXiv:2103.08153 [hep-ph]} \BibitemShut
  {NoStop}%
\bibitem [{\citenamefont {Choi}\ \emph {et~al.}(2021)\citenamefont {Choi},
  \citenamefont {Lin}, \citenamefont {Visinelli},\ and\ \citenamefont
  {Yanagida}}]{Choi:2021aze}%
  \BibitemOpen
  \bibfield  {author} {\bibinfo {author} {\bibfnamefont {G.}~\bibnamefont
  {Choi}}, \bibinfo {author} {\bibfnamefont {W.}~\bibnamefont {Lin}}, \bibinfo
  {author} {\bibfnamefont {L.}~\bibnamefont {Visinelli}}, \ and\ \bibinfo
  {author} {\bibfnamefont {T.~T.}\ \bibnamefont {Yanagida}},\ }\href {\doibase
  10.1103/PhysRevD.104.L101302} {\bibfield  {journal} {\bibinfo  {journal}
  {Phys. Rev. D}\ }\textbf {\bibinfo {volume} {104}},\ \bibinfo {pages}
  {L101302} (\bibinfo {year} {2021})},\ \Eprint
  {http://arxiv.org/abs/2106.12602} {arXiv:2106.12602 [hep-ph]} \BibitemShut
  {NoStop}%
\bibitem [{\citenamefont {Obata}(2022)}]{Obata:2021nql}%
  \BibitemOpen
  \bibfield  {author} {\bibinfo {author} {\bibfnamefont {I.}~\bibnamefont
  {Obata}},\ }\href {\doibase 10.1088/1475-7516/2022/09/062} {\bibfield
  {journal} {\bibinfo  {journal} {JCAP}\ }\textbf {\bibinfo {volume} {09}},\
  \bibinfo {pages} {062} (\bibinfo {year} {2022})},\ \Eprint
  {http://arxiv.org/abs/2108.02150} {arXiv:2108.02150 [astro-ph.CO]}
  \BibitemShut {NoStop}%
\bibitem [{\citenamefont {Yin}\ \emph {et~al.}(2022)\citenamefont {Yin},
  \citenamefont {Dai},\ and\ \citenamefont {Ferraro}}]{Yin:2021kmx}%
  \BibitemOpen
  \bibfield  {author} {\bibinfo {author} {\bibfnamefont {W.~W.}\ \bibnamefont
  {Yin}}, \bibinfo {author} {\bibfnamefont {L.}~\bibnamefont {Dai}}, \ and\
  \bibinfo {author} {\bibfnamefont {S.}~\bibnamefont {Ferraro}},\ }\href
  {\doibase 10.1088/1475-7516/2022/06/033} {\bibfield  {journal} {\bibinfo
  {journal} {JCAP}\ }\textbf {\bibinfo {volume} {06}},\ \bibinfo {pages} {033}
  (\bibinfo {year} {2022})},\ \Eprint {http://arxiv.org/abs/2111.12741}
  {arXiv:2111.12741 [astro-ph.CO]} \BibitemShut {NoStop}%
\bibitem [{\citenamefont {Kitajima}\ \emph {et~al.}(2022)\citenamefont
  {Kitajima}, \citenamefont {Kozai}, \citenamefont {Takahashi},\ and\
  \citenamefont {Yin}}]{Kitajima:2022jzz}%
  \BibitemOpen
  \bibfield  {author} {\bibinfo {author} {\bibfnamefont {N.}~\bibnamefont
  {Kitajima}}, \bibinfo {author} {\bibfnamefont {F.}~\bibnamefont {Kozai}},
  \bibinfo {author} {\bibfnamefont {F.}~\bibnamefont {Takahashi}}, \ and\
  \bibinfo {author} {\bibfnamefont {W.}~\bibnamefont {Yin}},\ }\href {\doibase
  10.1088/1475-7516/2022/10/043} {\bibfield  {journal} {\bibinfo  {journal}
  {JCAP}\ }\textbf {\bibinfo {volume} {10}},\ \bibinfo {pages} {043} (\bibinfo
  {year} {2022})},\ \Eprint {http://arxiv.org/abs/2205.05083} {arXiv:2205.05083
  [astro-ph.CO]} \BibitemShut {NoStop}%
\bibitem [{\citenamefont {Gasparotto}\ and\ \citenamefont
  {Obata}(2022)}]{Gasparotto:2022uqo}%
  \BibitemOpen
  \bibfield  {author} {\bibinfo {author} {\bibfnamefont {S.}~\bibnamefont
  {Gasparotto}}\ and\ \bibinfo {author} {\bibfnamefont {I.}~\bibnamefont
  {Obata}},\ }\href {\doibase 10.1088/1475-7516/2022/08/025} {\bibfield
  {journal} {\bibinfo  {journal} {JCAP}\ }\textbf {\bibinfo {volume} {08}},\
  \bibinfo {pages} {025} (\bibinfo {year} {2022})},\ \Eprint
  {http://arxiv.org/abs/2203.09409} {arXiv:2203.09409 [astro-ph.CO]}
  \BibitemShut {NoStop}%
\bibitem [{\citenamefont {Lin}\ and\ \citenamefont
  {Yanagida}(2023)}]{Lin:2022niw}%
  \BibitemOpen
  \bibfield  {author} {\bibinfo {author} {\bibfnamefont {W.}~\bibnamefont
  {Lin}}\ and\ \bibinfo {author} {\bibfnamefont {T.~T.}\ \bibnamefont
  {Yanagida}},\ }\href {\doibase 10.1103/PhysRevD.107.L021302} {\bibfield
  {journal} {\bibinfo  {journal} {Phys. Rev. D}\ }\textbf {\bibinfo {volume}
  {107}},\ \bibinfo {pages} {L021302} (\bibinfo {year} {2023})},\ \Eprint
  {http://arxiv.org/abs/2208.06843} {arXiv:2208.06843 [hep-ph]} \BibitemShut
  {NoStop}%
\bibitem [{\citenamefont {Murai}\ \emph {et~al.}(2023)\citenamefont {Murai},
  \citenamefont {Naokawa}, \citenamefont {Namikawa},\ and\ \citenamefont
  {Komatsu}}]{Murai:2022zur}%
  \BibitemOpen
  \bibfield  {author} {\bibinfo {author} {\bibfnamefont {K.}~\bibnamefont
  {Murai}}, \bibinfo {author} {\bibfnamefont {F.}~\bibnamefont {Naokawa}},
  \bibinfo {author} {\bibfnamefont {T.}~\bibnamefont {Namikawa}}, \ and\
  \bibinfo {author} {\bibfnamefont {E.}~\bibnamefont {Komatsu}},\ }\href
  {\doibase 10.1103/PhysRevD.107.L041302} {\bibfield  {journal} {\bibinfo
  {journal} {Phys. Rev. D}\ }\textbf {\bibinfo {volume} {107}},\ \bibinfo
  {pages} {L041302} (\bibinfo {year} {2023})},\ \Eprint
  {http://arxiv.org/abs/2209.07804} {arXiv:2209.07804 [astro-ph.CO]}
  \BibitemShut {NoStop}%
\bibitem [{\citenamefont {Gonzalez}\ \emph {et~al.}(2022)\citenamefont
  {Gonzalez}, \citenamefont {Kitajima}, \citenamefont {Takahashi},\ and\
  \citenamefont {Yin}}]{Gonzalez:2022mcx}%
  \BibitemOpen
  \bibfield  {author} {\bibinfo {author} {\bibfnamefont {D.}~\bibnamefont
  {Gonzalez}}, \bibinfo {author} {\bibfnamefont {N.}~\bibnamefont {Kitajima}},
  \bibinfo {author} {\bibfnamefont {F.}~\bibnamefont {Takahashi}}, \ and\
  \bibinfo {author} {\bibfnamefont {W.}~\bibnamefont {Yin}},\ }\href@noop {} {\
   (\bibinfo {year} {2022})},\ \Eprint {http://arxiv.org/abs/2211.06849}
  {arXiv:2211.06849 [hep-ph]} \BibitemShut {NoStop}%
\bibitem [{\citenamefont {Galaverni}\ \emph {et~al.}(2023)\citenamefont
  {Galaverni}, \citenamefont {Finelli},\ and\ \citenamefont
  {Paoletti}}]{Galaverni:2023zhv}%
  \BibitemOpen
  \bibfield  {author} {\bibinfo {author} {\bibfnamefont {M.}~\bibnamefont
  {Galaverni}}, \bibinfo {author} {\bibfnamefont {F.}~\bibnamefont {Finelli}},
  \ and\ \bibinfo {author} {\bibfnamefont {D.}~\bibnamefont {Paoletti}},\
  }\href {\doibase 10.1103/PhysRevD.107.083529} {\bibfield  {journal} {\bibinfo
   {journal} {Phys. Rev. D}\ }\textbf {\bibinfo {volume} {107}},\ \bibinfo
  {pages} {083529} (\bibinfo {year} {2023})},\ \Eprint
  {http://arxiv.org/abs/2301.07971} {arXiv:2301.07971 [astro-ph.CO]}
  \BibitemShut {NoStop}%
\bibitem [{\citenamefont {Minami}\ and\ \citenamefont
  {Komatsu}(2020)}]{Minami:2020odp}%
  \BibitemOpen
  \bibfield  {author} {\bibinfo {author} {\bibfnamefont {Y.}~\bibnamefont
  {Minami}}\ and\ \bibinfo {author} {\bibfnamefont {E.}~\bibnamefont
  {Komatsu}},\ }\href {\doibase 10.1103/PhysRevLett.125.221301} {\bibfield
  {journal} {\bibinfo  {journal} {Phys. Rev. Lett.}\ }\textbf {\bibinfo
  {volume} {125}},\ \bibinfo {pages} {221301} (\bibinfo {year} {2020})},\
  \Eprint {http://arxiv.org/abs/2011.11254} {arXiv:2011.11254 [astro-ph.CO]}
  \BibitemShut {NoStop}%
\bibitem [{\citenamefont {Diego-Palazuelos}\ \emph {et~al.}(2022)\citenamefont
  {Diego-Palazuelos} \emph {et~al.}}]{Diego-Palazuelos:2022dsq}%
  \BibitemOpen
  \bibfield  {author} {\bibinfo {author} {\bibfnamefont {P.}~\bibnamefont
  {Diego-Palazuelos}} \emph {et~al.},\ }\href {\doibase
  10.1103/PhysRevLett.128.091302} {\bibfield  {journal} {\bibinfo  {journal}
  {Phys. Rev. Lett.}\ }\textbf {\bibinfo {volume} {128}},\ \bibinfo {pages}
  {091302} (\bibinfo {year} {2022})},\ \Eprint
  {http://arxiv.org/abs/2201.07682} {arXiv:2201.07682 [astro-ph.CO]}
  \BibitemShut {NoStop}%
\bibitem [{\citenamefont {Eskilt}(2022)}]{Eskilt:2022wav}%
  \BibitemOpen
  \bibfield  {author} {\bibinfo {author} {\bibfnamefont {J.~R.}\ \bibnamefont
  {Eskilt}},\ }\href {\doibase 10.1051/0004-6361/202243269} {\bibfield
  {journal} {\bibinfo  {journal} {Astron. Astrophys.}\ }\textbf {\bibinfo
  {volume} {662}},\ \bibinfo {pages} {A10} (\bibinfo {year} {2022})},\ \Eprint
  {http://arxiv.org/abs/2201.13347} {arXiv:2201.13347 [astro-ph.CO]}
  \BibitemShut {NoStop}%
\bibitem [{\citenamefont {Eskilt}\ and\ \citenamefont
  {Komatsu}(2022)}]{Eskilt:2022cff}%
  \BibitemOpen
  \bibfield  {author} {\bibinfo {author} {\bibfnamefont {J.~R.}\ \bibnamefont
  {Eskilt}}\ and\ \bibinfo {author} {\bibfnamefont {E.}~\bibnamefont
  {Komatsu}},\ }\href {\doibase 10.1103/PhysRevD.106.063503} {\bibfield
  {journal} {\bibinfo  {journal} {Phys. Rev. D}\ }\textbf {\bibinfo {volume}
  {106}},\ \bibinfo {pages} {063503} (\bibinfo {year} {2022})},\ \Eprint
  {http://arxiv.org/abs/2205.13962} {arXiv:2205.13962 [astro-ph.CO]}
  \BibitemShut {NoStop}%
\bibitem [{\citenamefont {Balasubramanian}\ \emph {et~al.}(2005)\citenamefont
  {Balasubramanian}, \citenamefont {Berglund}, \citenamefont {Conlon},\ and\
  \citenamefont {Quevedo}}]{Balasubramanian:2005zx}%
  \BibitemOpen
  \bibfield  {author} {\bibinfo {author} {\bibfnamefont {V.}~\bibnamefont
  {Balasubramanian}}, \bibinfo {author} {\bibfnamefont {P.}~\bibnamefont
  {Berglund}}, \bibinfo {author} {\bibfnamefont {J.~P.}\ \bibnamefont
  {Conlon}}, \ and\ \bibinfo {author} {\bibfnamefont {F.}~\bibnamefont
  {Quevedo}},\ }\href {\doibase 10.1088/1126-6708/2005/03/007} {\bibfield
  {journal} {\bibinfo  {journal} {JHEP}\ }\textbf {\bibinfo {volume} {03}},\
  \bibinfo {pages} {007} (\bibinfo {year} {2005})},\ \Eprint
  {http://arxiv.org/abs/hep-th/0502058} {arXiv:hep-th/0502058} \BibitemShut
  {NoStop}%
\bibitem [{\citenamefont {Conlon}\ \emph {et~al.}(2005)\citenamefont {Conlon},
  \citenamefont {Quevedo},\ and\ \citenamefont {Suruliz}}]{Conlon:2005ki}%
  \BibitemOpen
  \bibfield  {author} {\bibinfo {author} {\bibfnamefont {J.~P.}\ \bibnamefont
  {Conlon}}, \bibinfo {author} {\bibfnamefont {F.}~\bibnamefont {Quevedo}}, \
  and\ \bibinfo {author} {\bibfnamefont {K.}~\bibnamefont {Suruliz}},\ }\href
  {\doibase 10.1088/1126-6708/2005/08/007} {\bibfield  {journal} {\bibinfo
  {journal} {JHEP}\ }\textbf {\bibinfo {volume} {08}},\ \bibinfo {pages} {007}
  (\bibinfo {year} {2005})},\ \Eprint {http://arxiv.org/abs/hep-th/0505076}
  {arXiv:hep-th/0505076} \BibitemShut {NoStop}%
\bibitem [{\citenamefont {Preskill}\ \emph {et~al.}(1983)\citenamefont
  {Preskill}, \citenamefont {Wise},\ and\ \citenamefont
  {Wilczek}}]{Preskill:1982cy}%
  \BibitemOpen
  \bibfield  {author} {\bibinfo {author} {\bibfnamefont {J.}~\bibnamefont
  {Preskill}}, \bibinfo {author} {\bibfnamefont {M.~B.}\ \bibnamefont {Wise}},
  \ and\ \bibinfo {author} {\bibfnamefont {F.}~\bibnamefont {Wilczek}},\ }\href
  {\doibase 10.1016/0370-2693(83)90637-8} {\bibfield  {journal} {\bibinfo
  {journal} {Phys. Lett. B}\ }\textbf {\bibinfo {volume} {120}},\ \bibinfo
  {pages} {127} (\bibinfo {year} {1983})}\BibitemShut {NoStop}%
\bibitem [{\citenamefont {Abbott}\ and\ \citenamefont
  {Sikivie}(1983)}]{Abbott:1982af}%
  \BibitemOpen
  \bibfield  {author} {\bibinfo {author} {\bibfnamefont {L.~F.}\ \bibnamefont
  {Abbott}}\ and\ \bibinfo {author} {\bibfnamefont {P.}~\bibnamefont
  {Sikivie}},\ }\href {\doibase 10.1016/0370-2693(83)90638-X} {\bibfield
  {journal} {\bibinfo  {journal} {Phys. Lett. B}\ }\textbf {\bibinfo {volume}
  {120}},\ \bibinfo {pages} {133} (\bibinfo {year} {1983})}\BibitemShut
  {NoStop}%
\bibitem [{\citenamefont {Dine}\ and\ \citenamefont
  {Fischler}(1983)}]{Dine:1982ah}%
  \BibitemOpen
  \bibfield  {author} {\bibinfo {author} {\bibfnamefont {M.}~\bibnamefont
  {Dine}}\ and\ \bibinfo {author} {\bibfnamefont {W.}~\bibnamefont
  {Fischler}},\ }\href {\doibase 10.1016/0370-2693(83)90639-1} {\bibfield
  {journal} {\bibinfo  {journal} {Phys. Lett. B}\ }\textbf {\bibinfo {volume}
  {120}},\ \bibinfo {pages} {137} (\bibinfo {year} {1983})}\BibitemShut
  {NoStop}%
\bibitem [{\citenamefont {de~Carlos}\ \emph {et~al.}(1993)\citenamefont
  {de~Carlos}, \citenamefont {Casas}, \citenamefont {Quevedo},\ and\
  \citenamefont {Roulet}}]{deCarlos:1993wie}%
  \BibitemOpen
  \bibfield  {author} {\bibinfo {author} {\bibfnamefont {B.}~\bibnamefont
  {de~Carlos}}, \bibinfo {author} {\bibfnamefont {J.~A.}\ \bibnamefont
  {Casas}}, \bibinfo {author} {\bibfnamefont {F.}~\bibnamefont {Quevedo}}, \
  and\ \bibinfo {author} {\bibfnamefont {E.}~\bibnamefont {Roulet}},\ }\href
  {\doibase 10.1016/0370-2693(93)91538-X} {\bibfield  {journal} {\bibinfo
  {journal} {Phys. Lett. B}\ }\textbf {\bibinfo {volume} {318}},\ \bibinfo
  {pages} {447} (\bibinfo {year} {1993})},\ \Eprint
  {http://arxiv.org/abs/hep-ph/9308325} {arXiv:hep-ph/9308325} \BibitemShut
  {NoStop}%
\bibitem [{\citenamefont {Banks}\ \emph {et~al.}(1994)\citenamefont {Banks},
  \citenamefont {Kaplan},\ and\ \citenamefont {Nelson}}]{Banks:1993en}%
  \BibitemOpen
  \bibfield  {author} {\bibinfo {author} {\bibfnamefont {T.}~\bibnamefont
  {Banks}}, \bibinfo {author} {\bibfnamefont {D.~B.}\ \bibnamefont {Kaplan}}, \
  and\ \bibinfo {author} {\bibfnamefont {A.~E.}\ \bibnamefont {Nelson}},\
  }\href {\doibase 10.1103/PhysRevD.49.779} {\bibfield  {journal} {\bibinfo
  {journal} {Phys. Rev. D}\ }\textbf {\bibinfo {volume} {49}},\ \bibinfo
  {pages} {779} (\bibinfo {year} {1994})},\ \Eprint
  {http://arxiv.org/abs/hep-ph/9308292} {arXiv:hep-ph/9308292} \BibitemShut
  {NoStop}%
\bibitem [{\citenamefont {Kitajima}\ \emph {et~al.}(2020)\citenamefont
  {Kitajima}, \citenamefont {Tada},\ and\ \citenamefont
  {Takahashi}}]{Kitajima:2019ibn}%
  \BibitemOpen
  \bibfield  {author} {\bibinfo {author} {\bibfnamefont {N.}~\bibnamefont
  {Kitajima}}, \bibinfo {author} {\bibfnamefont {Y.}~\bibnamefont {Tada}}, \
  and\ \bibinfo {author} {\bibfnamefont {F.}~\bibnamefont {Takahashi}},\ }\href
  {\doibase 10.1016/j.physletb.2019.135097} {\bibfield  {journal} {\bibinfo
  {journal} {Phys. Lett. B}\ }\textbf {\bibinfo {volume} {800}},\ \bibinfo
  {pages} {135097} (\bibinfo {year} {2020})},\ \Eprint
  {http://arxiv.org/abs/1908.08694} {arXiv:1908.08694 [hep-ph]} \BibitemShut
  {NoStop}%
\bibitem [{\citenamefont {Graham}\ and\ \citenamefont
  {Scherlis}(2018)}]{Graham:2018jyp}%
  \BibitemOpen
  \bibfield  {author} {\bibinfo {author} {\bibfnamefont {P.~W.}\ \bibnamefont
  {Graham}}\ and\ \bibinfo {author} {\bibfnamefont {A.}~\bibnamefont
  {Scherlis}},\ }\href {\doibase 10.1103/PhysRevD.98.035017} {\bibfield
  {journal} {\bibinfo  {journal} {Phys. Rev. D}\ }\textbf {\bibinfo {volume}
  {98}},\ \bibinfo {pages} {035017} (\bibinfo {year} {2018})},\ \Eprint
  {http://arxiv.org/abs/1805.07362} {arXiv:1805.07362 [hep-ph]} \BibitemShut
  {NoStop}%
\bibitem [{\citenamefont {Takahashi}\ \emph {et~al.}(2018)\citenamefont
  {Takahashi}, \citenamefont {Yin},\ and\ \citenamefont
  {Guth}}]{Takahashi:2018tdu}%
  \BibitemOpen
  \bibfield  {author} {\bibinfo {author} {\bibfnamefont {F.}~\bibnamefont
  {Takahashi}}, \bibinfo {author} {\bibfnamefont {W.}~\bibnamefont {Yin}}, \
  and\ \bibinfo {author} {\bibfnamefont {A.~H.}\ \bibnamefont {Guth}},\ }\href
  {\doibase 10.1103/PhysRevD.98.015042} {\bibfield  {journal} {\bibinfo
  {journal} {Phys. Rev. D}\ }\textbf {\bibinfo {volume} {98}},\ \bibinfo
  {pages} {015042} (\bibinfo {year} {2018})},\ \Eprint
  {http://arxiv.org/abs/1805.08763} {arXiv:1805.08763 [hep-ph]} \BibitemShut
  {NoStop}%
\bibitem [{\citenamefont {Ho}\ \emph {et~al.}(2019)\citenamefont {Ho},
  \citenamefont {Takahashi},\ and\ \citenamefont {Yin}}]{Ho:2019ayl}%
  \BibitemOpen
  \bibfield  {author} {\bibinfo {author} {\bibfnamefont {S.-Y.}\ \bibnamefont
  {Ho}}, \bibinfo {author} {\bibfnamefont {F.}~\bibnamefont {Takahashi}}, \
  and\ \bibinfo {author} {\bibfnamefont {W.}~\bibnamefont {Yin}},\ }\href
  {\doibase 10.1007/JHEP04(2019)149} {\bibfield  {journal} {\bibinfo  {journal}
  {JHEP}\ }\textbf {\bibinfo {volume} {04}},\ \bibinfo {pages} {149} (\bibinfo
  {year} {2019})},\ \Eprint {http://arxiv.org/abs/1901.01240} {arXiv:1901.01240
  [hep-ph]} \BibitemShut {NoStop}%
\bibitem [{\citenamefont {Reig}(2021)}]{Reig:2021ipa}%
  \BibitemOpen
  \bibfield  {author} {\bibinfo {author} {\bibfnamefont {M.}~\bibnamefont
  {Reig}},\ }\href {\doibase 10.1007/JHEP09(2021)207} {\bibfield  {journal}
  {\bibinfo  {journal} {JHEP}\ }\textbf {\bibinfo {volume} {09}},\ \bibinfo
  {pages} {207} (\bibinfo {year} {2021})},\ \Eprint
  {http://arxiv.org/abs/2104.09923} {arXiv:2104.09923 [hep-ph]} \BibitemShut
  {NoStop}%
\bibitem [{\citenamefont {Alonso-\'Alvarez}\ \emph {et~al.}(2020)\citenamefont
  {Alonso-\'Alvarez}, \citenamefont {Hugle},\ and\ \citenamefont
  {Jaeckel}}]{Alonso-Alvarez:2019ixv}%
  \BibitemOpen
  \bibfield  {author} {\bibinfo {author} {\bibfnamefont {G.}~\bibnamefont
  {Alonso-\'Alvarez}}, \bibinfo {author} {\bibfnamefont {T.}~\bibnamefont
  {Hugle}}, \ and\ \bibinfo {author} {\bibfnamefont {J.}~\bibnamefont
  {Jaeckel}},\ }\href {\doibase 10.1088/1475-7516/2020/02/014} {\bibfield
  {journal} {\bibinfo  {journal} {JCAP}\ }\textbf {\bibinfo {volume} {02}},\
  \bibinfo {pages} {014} (\bibinfo {year} {2020})},\ \Eprint
  {http://arxiv.org/abs/1905.09836} {arXiv:1905.09836 [hep-ph]} \BibitemShut
  {NoStop}%
\bibitem [{\citenamefont {Daido}\ \emph {et~al.}(2017)\citenamefont {Daido},
  \citenamefont {Kobayashi},\ and\ \citenamefont {Takahashi}}]{Daido:2016tsj}%
  \BibitemOpen
  \bibfield  {author} {\bibinfo {author} {\bibfnamefont {R.}~\bibnamefont
  {Daido}}, \bibinfo {author} {\bibfnamefont {T.}~\bibnamefont {Kobayashi}}, \
  and\ \bibinfo {author} {\bibfnamefont {F.}~\bibnamefont {Takahashi}},\ }\href
  {\doibase 10.1016/j.physletb.2016.12.034} {\bibfield  {journal} {\bibinfo
  {journal} {Phys. Lett. B}\ }\textbf {\bibinfo {volume} {765}},\ \bibinfo
  {pages} {293} (\bibinfo {year} {2017})},\ \Eprint
  {http://arxiv.org/abs/1608.04092} {arXiv:1608.04092 [hep-ph]} \BibitemShut
  {NoStop}%
\bibitem [{\citenamefont {Nakagawa}\ \emph {et~al.}(2020)\citenamefont
  {Nakagawa}, \citenamefont {Takahashi},\ and\ \citenamefont
  {Yin}}]{Nakagawa:2020eeg}%
  \BibitemOpen
  \bibfield  {author} {\bibinfo {author} {\bibfnamefont {S.}~\bibnamefont
  {Nakagawa}}, \bibinfo {author} {\bibfnamefont {F.}~\bibnamefont {Takahashi}},
  \ and\ \bibinfo {author} {\bibfnamefont {W.}~\bibnamefont {Yin}},\ }\href
  {\doibase 10.1088/1475-7516/2020/05/004} {\bibfield  {journal} {\bibinfo
  {journal} {JCAP}\ }\textbf {\bibinfo {volume} {05}},\ \bibinfo {pages} {004}
  (\bibinfo {year} {2020})},\ \Eprint {http://arxiv.org/abs/2002.12195}
  {arXiv:2002.12195 [hep-ph]} \BibitemShut {NoStop}%
\bibitem [{\citenamefont {Matsui}\ \emph {et~al.}(2020)\citenamefont {Matsui},
  \citenamefont {Takahashi},\ and\ \citenamefont {Yin}}]{Matsui:2020wfx}%
  \BibitemOpen
  \bibfield  {author} {\bibinfo {author} {\bibfnamefont {H.}~\bibnamefont
  {Matsui}}, \bibinfo {author} {\bibfnamefont {F.}~\bibnamefont {Takahashi}}, \
  and\ \bibinfo {author} {\bibfnamefont {W.}~\bibnamefont {Yin}},\ }\href
  {\doibase 10.1007/JHEP05(2020)154} {\bibfield  {journal} {\bibinfo  {journal}
  {JHEP}\ }\textbf {\bibinfo {volume} {05}},\ \bibinfo {pages} {154} (\bibinfo
  {year} {2020})},\ \Eprint {http://arxiv.org/abs/2001.04464} {arXiv:2001.04464
  [hep-ph]} \BibitemShut {NoStop}%
\bibitem [{\citenamefont {Takahashi}\ and\ \citenamefont
  {Yin}(2019{\natexlab{a}})}]{Takahashi:2019qmh}%
  \BibitemOpen
  \bibfield  {author} {\bibinfo {author} {\bibfnamefont {F.}~\bibnamefont
  {Takahashi}}\ and\ \bibinfo {author} {\bibfnamefont {W.}~\bibnamefont
  {Yin}},\ }\href {\doibase 10.1007/JHEP07(2019)095} {\bibfield  {journal}
  {\bibinfo  {journal} {JHEP}\ }\textbf {\bibinfo {volume} {07}},\ \bibinfo
  {pages} {095} (\bibinfo {year} {2019}{\natexlab{a}})},\ \Eprint
  {http://arxiv.org/abs/1903.00462} {arXiv:1903.00462 [hep-ph]} \BibitemShut
  {NoStop}%
\bibitem [{\citenamefont {Takahashi}\ and\ \citenamefont
  {Yin}(2019{\natexlab{b}})}]{Takahashi:2019pqf}%
  \BibitemOpen
  \bibfield  {author} {\bibinfo {author} {\bibfnamefont {F.}~\bibnamefont
  {Takahashi}}\ and\ \bibinfo {author} {\bibfnamefont {W.}~\bibnamefont
  {Yin}},\ }\href {\doibase 10.1007/JHEP10(2019)120} {\bibfield  {journal}
  {\bibinfo  {journal} {JHEP}\ }\textbf {\bibinfo {volume} {10}},\ \bibinfo
  {pages} {120} (\bibinfo {year} {2019}{\natexlab{b}})},\ \Eprint
  {http://arxiv.org/abs/1908.06071} {arXiv:1908.06071 [hep-ph]} \BibitemShut
  {NoStop}%
\bibitem [{\citenamefont {Kim}\ \emph {et~al.}(2005)\citenamefont {Kim},
  \citenamefont {Nilles},\ and\ \citenamefont {Peloso}}]{Kim:2004rp}%
  \BibitemOpen
  \bibfield  {author} {\bibinfo {author} {\bibfnamefont {J.~E.}\ \bibnamefont
  {Kim}}, \bibinfo {author} {\bibfnamefont {H.~P.}\ \bibnamefont {Nilles}}, \
  and\ \bibinfo {author} {\bibfnamefont {M.}~\bibnamefont {Peloso}},\ }\href
  {\doibase 10.1088/1475-7516/2005/01/005} {\bibfield  {journal} {\bibinfo
  {journal} {JCAP}\ }\textbf {\bibinfo {volume} {01}},\ \bibinfo {pages} {005}
  (\bibinfo {year} {2005})},\ \Eprint {http://arxiv.org/abs/hep-ph/0409138}
  {arXiv:hep-ph/0409138} \BibitemShut {NoStop}%
\bibitem [{\citenamefont {Choi}\ \emph {et~al.}(2014)\citenamefont {Choi},
  \citenamefont {Kim},\ and\ \citenamefont {Yun}}]{Choi:2014rja}%
  \BibitemOpen
  \bibfield  {author} {\bibinfo {author} {\bibfnamefont {K.}~\bibnamefont
  {Choi}}, \bibinfo {author} {\bibfnamefont {H.}~\bibnamefont {Kim}}, \ and\
  \bibinfo {author} {\bibfnamefont {S.}~\bibnamefont {Yun}},\ }\href {\doibase
  10.1103/PhysRevD.90.023545} {\bibfield  {journal} {\bibinfo  {journal} {Phys.
  Rev. D}\ }\textbf {\bibinfo {volume} {90}},\ \bibinfo {pages} {023545}
  (\bibinfo {year} {2014})},\ \Eprint {http://arxiv.org/abs/1404.6209}
  {arXiv:1404.6209 [hep-th]} \BibitemShut {NoStop}%
\bibitem [{\citenamefont {Higaki}\ and\ \citenamefont
  {Takahashi}(2015)}]{Higaki:2014mwa}%
  \BibitemOpen
  \bibfield  {author} {\bibinfo {author} {\bibfnamefont {T.}~\bibnamefont
  {Higaki}}\ and\ \bibinfo {author} {\bibfnamefont {F.}~\bibnamefont
  {Takahashi}},\ }\href {\doibase 10.1016/j.physletb.2015.03.052} {\bibfield
  {journal} {\bibinfo  {journal} {Phys. Lett. B}\ }\textbf {\bibinfo {volume}
  {744}},\ \bibinfo {pages} {153} (\bibinfo {year} {2015})},\ \Eprint
  {http://arxiv.org/abs/1409.8409} {arXiv:1409.8409 [hep-ph]} \BibitemShut
  {NoStop}%
\bibitem [{\citenamefont {Kaplan}\ and\ \citenamefont
  {Rattazzi}(2016)}]{Kaplan:2015fuy}%
  \BibitemOpen
  \bibfield  {author} {\bibinfo {author} {\bibfnamefont {D.~E.}\ \bibnamefont
  {Kaplan}}\ and\ \bibinfo {author} {\bibfnamefont {R.}~\bibnamefont
  {Rattazzi}},\ }\href {\doibase 10.1103/PhysRevD.93.085007} {\bibfield
  {journal} {\bibinfo  {journal} {Phys. Rev. D}\ }\textbf {\bibinfo {volume}
  {93}},\ \bibinfo {pages} {085007} (\bibinfo {year} {2016})},\ \Eprint
  {http://arxiv.org/abs/1511.01827} {arXiv:1511.01827 [hep-ph]} \BibitemShut
  {NoStop}%
\bibitem [{\citenamefont {Giudice}\ and\ \citenamefont
  {McCullough}(2017)}]{Giudice:2016yja}%
  \BibitemOpen
  \bibfield  {author} {\bibinfo {author} {\bibfnamefont {G.~F.}\ \bibnamefont
  {Giudice}}\ and\ \bibinfo {author} {\bibfnamefont {M.}~\bibnamefont
  {McCullough}},\ }\href {\doibase 10.1007/JHEP02(2017)036} {\bibfield
  {journal} {\bibinfo  {journal} {JHEP}\ }\textbf {\bibinfo {volume} {02}},\
  \bibinfo {pages} {036} (\bibinfo {year} {2017})},\ \Eprint
  {http://arxiv.org/abs/1610.07962} {arXiv:1610.07962 [hep-ph]} \BibitemShut
  {NoStop}%
\bibitem [{\citenamefont {Higaki}\ \emph
  {et~al.}(2016{\natexlab{a}})\citenamefont {Higaki}, \citenamefont {Jeong},
  \citenamefont {Kitajima}, \citenamefont {Sekiguchi},\ and\ \citenamefont
  {Takahashi}}]{Higaki:2016jjh}%
  \BibitemOpen
  \bibfield  {author} {\bibinfo {author} {\bibfnamefont {T.}~\bibnamefont
  {Higaki}}, \bibinfo {author} {\bibfnamefont {K.~S.}\ \bibnamefont {Jeong}},
  \bibinfo {author} {\bibfnamefont {N.}~\bibnamefont {Kitajima}}, \bibinfo
  {author} {\bibfnamefont {T.}~\bibnamefont {Sekiguchi}}, \ and\ \bibinfo
  {author} {\bibfnamefont {F.}~\bibnamefont {Takahashi}},\ }\href {\doibase
  10.1007/JHEP08(2016)044} {\bibfield  {journal} {\bibinfo  {journal} {JHEP}\
  }\textbf {\bibinfo {volume} {08}},\ \bibinfo {pages} {044} (\bibinfo {year}
  {2016}{\natexlab{a}})},\ \Eprint {http://arxiv.org/abs/1606.05552}
  {arXiv:1606.05552 [hep-ph]} \BibitemShut {NoStop}%
\bibitem [{\citenamefont {Higaki}\ \emph
  {et~al.}(2016{\natexlab{b}})\citenamefont {Higaki}, \citenamefont {Jeong},
  \citenamefont {Kitajima},\ and\ \citenamefont {Takahashi}}]{Higaki:2016yqk}%
  \BibitemOpen
  \bibfield  {author} {\bibinfo {author} {\bibfnamefont {T.}~\bibnamefont
  {Higaki}}, \bibinfo {author} {\bibfnamefont {K.~S.}\ \bibnamefont {Jeong}},
  \bibinfo {author} {\bibfnamefont {N.}~\bibnamefont {Kitajima}}, \ and\
  \bibinfo {author} {\bibfnamefont {F.}~\bibnamefont {Takahashi}},\ }\href
  {\doibase 10.1007/JHEP06(2016)150} {\bibfield  {journal} {\bibinfo  {journal}
  {JHEP}\ }\textbf {\bibinfo {volume} {06}},\ \bibinfo {pages} {150} (\bibinfo
  {year} {2016}{\natexlab{b}})},\ \Eprint {http://arxiv.org/abs/1603.02090}
  {arXiv:1603.02090 [hep-ph]} \BibitemShut {NoStop}%
\bibitem [{\citenamefont {Gavela}\ \emph {et~al.}(2023)\citenamefont {Gavela},
  \citenamefont {Qu\'\i{}lez},\ and\ \citenamefont {Ramos}}]{Gavela:2023tzu}%
  \BibitemOpen
  \bibfield  {author} {\bibinfo {author} {\bibfnamefont {B.}~\bibnamefont
  {Gavela}}, \bibinfo {author} {\bibfnamefont {P.}~\bibnamefont {Qu\'\i{}lez}},
  \ and\ \bibinfo {author} {\bibfnamefont {M.}~\bibnamefont {Ramos}},\
  }\href@noop {} {\  (\bibinfo {year} {2023})},\ \Eprint
  {http://arxiv.org/abs/2305.15465} {arXiv:2305.15465 [hep-ph]} \BibitemShut
  {NoStop}%
\bibitem [{\citenamefont {Kitajima}\ and\ \citenamefont
  {Takahashi}(2015)}]{Kitajima:2014xla}%
  \BibitemOpen
  \bibfield  {author} {\bibinfo {author} {\bibfnamefont {N.}~\bibnamefont
  {Kitajima}}\ and\ \bibinfo {author} {\bibfnamefont {F.}~\bibnamefont
  {Takahashi}},\ }\href {\doibase 10.1088/1475-7516/2015/01/032} {\bibfield
  {journal} {\bibinfo  {journal} {JCAP}\ }\textbf {\bibinfo {volume} {01}},\
  \bibinfo {pages} {032} (\bibinfo {year} {2015})},\ \Eprint
  {http://arxiv.org/abs/1411.2011} {arXiv:1411.2011 [hep-ph]} \BibitemShut
  {NoStop}%
\bibitem [{\citenamefont {Daido}\ \emph {et~al.}(2016)\citenamefont {Daido},
  \citenamefont {Kitajima},\ and\ \citenamefont {Takahashi}}]{Daido:2015cba}%
  \BibitemOpen
  \bibfield  {author} {\bibinfo {author} {\bibfnamefont {R.}~\bibnamefont
  {Daido}}, \bibinfo {author} {\bibfnamefont {N.}~\bibnamefont {Kitajima}}, \
  and\ \bibinfo {author} {\bibfnamefont {F.}~\bibnamefont {Takahashi}},\ }\href
  {\doibase 10.1103/PhysRevD.93.075027} {\bibfield  {journal} {\bibinfo
  {journal} {Phys. Rev. D}\ }\textbf {\bibinfo {volume} {93}},\ \bibinfo
  {pages} {075027} (\bibinfo {year} {2016})},\ \Eprint
  {http://arxiv.org/abs/1510.06675} {arXiv:1510.06675 [hep-ph]} \BibitemShut
  {NoStop}%
\bibitem [{\citenamefont {Ho}\ \emph {et~al.}(2018)\citenamefont {Ho},
  \citenamefont {Saikawa},\ and\ \citenamefont {Takahashi}}]{Ho:2018qur}%
  \BibitemOpen
  \bibfield  {author} {\bibinfo {author} {\bibfnamefont {S.-Y.}\ \bibnamefont
  {Ho}}, \bibinfo {author} {\bibfnamefont {K.}~\bibnamefont {Saikawa}}, \ and\
  \bibinfo {author} {\bibfnamefont {F.}~\bibnamefont {Takahashi}},\ }\href
  {\doibase 10.1088/1475-7516/2018/10/042} {\bibfield  {journal} {\bibinfo
  {journal} {JCAP}\ }\textbf {\bibinfo {volume} {10}},\ \bibinfo {pages} {042}
  (\bibinfo {year} {2018})},\ \Eprint {http://arxiv.org/abs/1806.09551}
  {arXiv:1806.09551 [hep-ph]} \BibitemShut {NoStop}%
\bibitem [{\citenamefont {Daido}\ \emph {et~al.}(2015)\citenamefont {Daido},
  \citenamefont {Kitajima},\ and\ \citenamefont {Takahashi}}]{Daido:2015bva}%
  \BibitemOpen
  \bibfield  {author} {\bibinfo {author} {\bibfnamefont {R.}~\bibnamefont
  {Daido}}, \bibinfo {author} {\bibfnamefont {N.}~\bibnamefont {Kitajima}}, \
  and\ \bibinfo {author} {\bibfnamefont {F.}~\bibnamefont {Takahashi}},\ }\href
  {\doibase 10.1103/PhysRevD.92.063512} {\bibfield  {journal} {\bibinfo
  {journal} {Phys. Rev. D}\ }\textbf {\bibinfo {volume} {92}},\ \bibinfo
  {pages} {063512} (\bibinfo {year} {2015})},\ \Eprint
  {http://arxiv.org/abs/1505.07670} {arXiv:1505.07670 [hep-ph]} \BibitemShut
  {NoStop}%
\bibitem [{\citenamefont {Arias}\ \emph {et~al.}(2012)\citenamefont {Arias},
  \citenamefont {Cadamuro}, \citenamefont {Goodsell}, \citenamefont {Jaeckel},
  \citenamefont {Redondo},\ and\ \citenamefont {Ringwald}}]{Arias:2012az}%
  \BibitemOpen
  \bibfield  {author} {\bibinfo {author} {\bibfnamefont {P.}~\bibnamefont
  {Arias}}, \bibinfo {author} {\bibfnamefont {D.}~\bibnamefont {Cadamuro}},
  \bibinfo {author} {\bibfnamefont {M.}~\bibnamefont {Goodsell}}, \bibinfo
  {author} {\bibfnamefont {J.}~\bibnamefont {Jaeckel}}, \bibinfo {author}
  {\bibfnamefont {J.}~\bibnamefont {Redondo}}, \ and\ \bibinfo {author}
  {\bibfnamefont {A.}~\bibnamefont {Ringwald}},\ }\href {\doibase
  10.1088/1475-7516/2012/06/013} {\bibfield  {journal} {\bibinfo  {journal}
  {JCAP}\ }\textbf {\bibinfo {volume} {06}},\ \bibinfo {pages} {013} (\bibinfo
  {year} {2012})},\ \Eprint {http://arxiv.org/abs/1201.5902} {arXiv:1201.5902
  [hep-ph]} \BibitemShut {NoStop}%
\bibitem [{\citenamefont {Nakagawa}\ \emph {et~al.}(2023)\citenamefont
  {Nakagawa}, \citenamefont {Takahashi}, \citenamefont {Yamada},\ and\
  \citenamefont {Yin}}]{Nakagawa:2022wwm}%
  \BibitemOpen
  \bibfield  {author} {\bibinfo {author} {\bibfnamefont {S.}~\bibnamefont
  {Nakagawa}}, \bibinfo {author} {\bibfnamefont {F.}~\bibnamefont {Takahashi}},
  \bibinfo {author} {\bibfnamefont {M.}~\bibnamefont {Yamada}}, \ and\ \bibinfo
  {author} {\bibfnamefont {W.}~\bibnamefont {Yin}},\ }\href {\doibase
  10.1016/j.physletb.2023.137824} {\bibfield  {journal} {\bibinfo  {journal}
  {Phys. Lett. B}\ }\textbf {\bibinfo {volume} {839}},\ \bibinfo {pages}
  {137824} (\bibinfo {year} {2023})},\ \Eprint
  {http://arxiv.org/abs/2210.10022} {arXiv:2210.10022 [hep-ph]} \BibitemShut
  {NoStop}%
\bibitem [{\citenamefont {Borsanyi}\ \emph {et~al.}(2016)\citenamefont
  {Borsanyi} \emph {et~al.}}]{Borsanyi:2016ksw}%
  \BibitemOpen
  \bibfield  {author} {\bibinfo {author} {\bibfnamefont {S.}~\bibnamefont
  {Borsanyi}} \emph {et~al.},\ }\href {\doibase 10.1038/nature20115} {\bibfield
   {journal} {\bibinfo  {journal} {Nature}\ }\textbf {\bibinfo {volume}
  {539}},\ \bibinfo {pages} {69} (\bibinfo {year} {2016})},\ \Eprint
  {http://arxiv.org/abs/1606.07494} {arXiv:1606.07494 [hep-lat]} \BibitemShut
  {NoStop}%
\bibitem [{\citenamefont {Gibbons}\ and\ \citenamefont
  {Hawking}(1977)}]{Gibbons:1977mu}%
  \BibitemOpen
  \bibfield  {author} {\bibinfo {author} {\bibfnamefont {G.~W.}\ \bibnamefont
  {Gibbons}}\ and\ \bibinfo {author} {\bibfnamefont {S.~W.}\ \bibnamefont
  {Hawking}},\ }\href {\doibase 10.1103/PhysRevD.15.2738} {\bibfield  {journal}
  {\bibinfo  {journal} {Phys. Rev. D}\ }\textbf {\bibinfo {volume} {15}},\
  \bibinfo {pages} {2738} (\bibinfo {year} {1977})}\BibitemShut {NoStop}%
\bibitem [{\citenamefont {Barenboim}\ \emph {et~al.}(2016)\citenamefont
  {Barenboim}, \citenamefont {Park},\ and\ \citenamefont
  {Kinney}}]{Barenboim:2016mmw}%
  \BibitemOpen
  \bibfield  {author} {\bibinfo {author} {\bibfnamefont {G.}~\bibnamefont
  {Barenboim}}, \bibinfo {author} {\bibfnamefont {W.-I.}\ \bibnamefont {Park}},
  \ and\ \bibinfo {author} {\bibfnamefont {W.~H.}\ \bibnamefont {Kinney}},\
  }\href {\doibase 10.1088/1475-7516/2016/05/030} {\bibfield  {journal}
  {\bibinfo  {journal} {JCAP}\ }\textbf {\bibinfo {volume} {05}},\ \bibinfo
  {pages} {030} (\bibinfo {year} {2016})},\ \Eprint
  {http://arxiv.org/abs/1601.08140} {arXiv:1601.08140 [astro-ph.CO]}
  \BibitemShut {NoStop}%
\bibitem [{\citenamefont {Assadullahi}\ \emph {et~al.}(2016)\citenamefont
  {Assadullahi}, \citenamefont {Firouzjahi}, \citenamefont {Noorbala},
  \citenamefont {Vennin},\ and\ \citenamefont {Wands}}]{Assadullahi:2016gkk}%
  \BibitemOpen
  \bibfield  {author} {\bibinfo {author} {\bibfnamefont {H.}~\bibnamefont
  {Assadullahi}}, \bibinfo {author} {\bibfnamefont {H.}~\bibnamefont
  {Firouzjahi}}, \bibinfo {author} {\bibfnamefont {M.}~\bibnamefont
  {Noorbala}}, \bibinfo {author} {\bibfnamefont {V.}~\bibnamefont {Vennin}}, \
  and\ \bibinfo {author} {\bibfnamefont {D.}~\bibnamefont {Wands}},\ }\href
  {\doibase 10.1088/1475-7516/2016/06/043} {\bibfield  {journal} {\bibinfo
  {journal} {JCAP}\ }\textbf {\bibinfo {volume} {06}},\ \bibinfo {pages} {043}
  (\bibinfo {year} {2016})},\ \Eprint {http://arxiv.org/abs/1604.04502}
  {arXiv:1604.04502 [hep-th]} \BibitemShut {NoStop}%
\bibitem [{\citenamefont {Dubovsky}\ \emph {et~al.}(2012)\citenamefont
  {Dubovsky}, \citenamefont {Senatore},\ and\ \citenamefont
  {Villadoro}}]{Dubovsky:2011uy}%
  \BibitemOpen
  \bibfield  {author} {\bibinfo {author} {\bibfnamefont {S.}~\bibnamefont
  {Dubovsky}}, \bibinfo {author} {\bibfnamefont {L.}~\bibnamefont {Senatore}},
  \ and\ \bibinfo {author} {\bibfnamefont {G.}~\bibnamefont {Villadoro}},\
  }\href {\doibase 10.1007/JHEP05(2012)035} {\bibfield  {journal} {\bibinfo
  {journal} {JHEP}\ }\textbf {\bibinfo {volume} {05}},\ \bibinfo {pages} {035}
  (\bibinfo {year} {2012})},\ \Eprint {http://arxiv.org/abs/1111.1725}
  {arXiv:1111.1725 [hep-th]} \BibitemShut {NoStop}%
\bibitem [{\citenamefont {Starobinsky}(1986)}]{Starobinsky:1986fx}%
  \BibitemOpen
  \bibfield  {author} {\bibinfo {author} {\bibfnamefont {A.~A.}\ \bibnamefont
  {Starobinsky}},\ }\href {\doibase 10.1007/3-540-16452-9_6} {\bibfield
  {journal} {\bibinfo  {journal} {Lect. Notes Phys.}\ }\textbf {\bibinfo
  {volume} {246}},\ \bibinfo {pages} {107} (\bibinfo {year}
  {1986})}\BibitemShut {NoStop}%
\bibitem [{\citenamefont {Starobinsky}\ and\ \citenamefont
  {Yokoyama}(1994)}]{Starobinsky:1994bd}%
  \BibitemOpen
  \bibfield  {author} {\bibinfo {author} {\bibfnamefont {A.~A.}\ \bibnamefont
  {Starobinsky}}\ and\ \bibinfo {author} {\bibfnamefont {J.}~\bibnamefont
  {Yokoyama}},\ }\href {\doibase 10.1103/PhysRevD.50.6357} {\bibfield
  {journal} {\bibinfo  {journal} {Phys. Rev. D}\ }\textbf {\bibinfo {volume}
  {50}},\ \bibinfo {pages} {6357} (\bibinfo {year} {1994})},\ \Eprint
  {http://arxiv.org/abs/astro-ph/9407016} {arXiv:astro-ph/9407016} \BibitemShut
  {NoStop}%
\bibitem [{\citenamefont {Nakao}\ \emph {et~al.}(1988)\citenamefont {Nakao},
  \citenamefont {Nambu},\ and\ \citenamefont {Sasaki}}]{Nakao:1988yi}%
  \BibitemOpen
  \bibfield  {author} {\bibinfo {author} {\bibfnamefont {K.-i.}\ \bibnamefont
  {Nakao}}, \bibinfo {author} {\bibfnamefont {Y.}~\bibnamefont {Nambu}}, \ and\
  \bibinfo {author} {\bibfnamefont {M.}~\bibnamefont {Sasaki}},\ }\href
  {\doibase 10.1143/PTP.80.1041} {\bibfield  {journal} {\bibinfo  {journal}
  {Prog. Theor. Phys.}\ }\textbf {\bibinfo {volume} {80}},\ \bibinfo {pages}
  {1041} (\bibinfo {year} {1988})}\BibitemShut {NoStop}%
\bibitem [{\citenamefont {Nambu}\ and\ \citenamefont
  {Sasaki}(1989)}]{Nambu:1988je}%
  \BibitemOpen
  \bibfield  {author} {\bibinfo {author} {\bibfnamefont {Y.}~\bibnamefont
  {Nambu}}\ and\ \bibinfo {author} {\bibfnamefont {M.}~\bibnamefont {Sasaki}},\
  }\href {\doibase 10.1016/0370-2693(89)90385-7} {\bibfield  {journal}
  {\bibinfo  {journal} {Phys. Lett. B}\ }\textbf {\bibinfo {volume} {219}},\
  \bibinfo {pages} {240} (\bibinfo {year} {1989})}\BibitemShut {NoStop}%
\bibitem [{\citenamefont {Nambu}(1989)}]{Nambu:1989uf}%
  \BibitemOpen
  \bibfield  {author} {\bibinfo {author} {\bibfnamefont {Y.}~\bibnamefont
  {Nambu}},\ }\href {\doibase 10.1143/PTP.81.1037} {\bibfield  {journal}
  {\bibinfo  {journal} {Prog. Theor. Phys.}\ }\textbf {\bibinfo {volume}
  {81}},\ \bibinfo {pages} {1037} (\bibinfo {year} {1989})}\BibitemShut
  {NoStop}%
\bibitem [{\citenamefont {Linde}\ \emph {et~al.}(1994)\citenamefont {Linde},
  \citenamefont {Linde},\ and\ \citenamefont {Mezhlumian}}]{Linde:1993xx}%
  \BibitemOpen
  \bibfield  {author} {\bibinfo {author} {\bibfnamefont {A.~D.}\ \bibnamefont
  {Linde}}, \bibinfo {author} {\bibfnamefont {D.~A.}\ \bibnamefont {Linde}}, \
  and\ \bibinfo {author} {\bibfnamefont {A.}~\bibnamefont {Mezhlumian}},\
  }\href {\doibase 10.1103/PhysRevD.49.1783} {\bibfield  {journal} {\bibinfo
  {journal} {Phys. Rev. D}\ }\textbf {\bibinfo {volume} {49}},\ \bibinfo
  {pages} {1783} (\bibinfo {year} {1994})},\ \Eprint
  {http://arxiv.org/abs/gr-qc/9306035} {arXiv:gr-qc/9306035} \BibitemShut
  {NoStop}%
\bibitem [{\citenamefont {Bae}\ \emph {et~al.}(2008)\citenamefont {Bae},
  \citenamefont {Huh},\ and\ \citenamefont {Kim}}]{Bae:2008ue}%
  \BibitemOpen
  \bibfield  {author} {\bibinfo {author} {\bibfnamefont {K.~J.}\ \bibnamefont
  {Bae}}, \bibinfo {author} {\bibfnamefont {J.-H.}\ \bibnamefont {Huh}}, \ and\
  \bibinfo {author} {\bibfnamefont {J.~E.}\ \bibnamefont {Kim}},\ }\href
  {\doibase 10.1088/1475-7516/2008/09/005} {\bibfield  {journal} {\bibinfo
  {journal} {JCAP}\ }\textbf {\bibinfo {volume} {09}},\ \bibinfo {pages} {005}
  (\bibinfo {year} {2008})},\ \Eprint {http://arxiv.org/abs/0806.0497}
  {arXiv:0806.0497 [hep-ph]} \BibitemShut {NoStop}%
\bibitem [{\citenamefont {Visinelli}\ and\ \citenamefont
  {Gondolo}(2009)}]{Visinelli:2009zm}%
  \BibitemOpen
  \bibfield  {author} {\bibinfo {author} {\bibfnamefont {L.}~\bibnamefont
  {Visinelli}}\ and\ \bibinfo {author} {\bibfnamefont {P.}~\bibnamefont
  {Gondolo}},\ }\href {\doibase 10.1103/PhysRevD.80.035024} {\bibfield
  {journal} {\bibinfo  {journal} {Phys. Rev. D}\ }\textbf {\bibinfo {volume}
  {80}},\ \bibinfo {pages} {035024} (\bibinfo {year} {2009})},\ \Eprint
  {http://arxiv.org/abs/0903.4377} {arXiv:0903.4377 [astro-ph.CO]} \BibitemShut
  {NoStop}%
\bibitem [{\citenamefont {Ballesteros}\ \emph {et~al.}(2017)\citenamefont
  {Ballesteros}, \citenamefont {Redondo}, \citenamefont {Ringwald},\ and\
  \citenamefont {Tamarit}}]{Ballesteros:2016xej}%
  \BibitemOpen
  \bibfield  {author} {\bibinfo {author} {\bibfnamefont {G.}~\bibnamefont
  {Ballesteros}}, \bibinfo {author} {\bibfnamefont {J.}~\bibnamefont
  {Redondo}}, \bibinfo {author} {\bibfnamefont {A.}~\bibnamefont {Ringwald}}, \
  and\ \bibinfo {author} {\bibfnamefont {C.}~\bibnamefont {Tamarit}},\ }\href
  {\doibase 10.1088/1475-7516/2017/08/001} {\bibfield  {journal} {\bibinfo
  {journal} {JCAP}\ }\textbf {\bibinfo {volume} {08}},\ \bibinfo {pages} {001}
  (\bibinfo {year} {2017})},\ \Eprint {http://arxiv.org/abs/1610.01639}
  {arXiv:1610.01639 [hep-ph]} \BibitemShut {NoStop}%
\bibitem [{\citenamefont {Silva-Feaver}\ \emph {et~al.}(2017)\citenamefont
  {Silva-Feaver} \emph {et~al.}}]{Silva-Feaver:2016qhh}%
  \BibitemOpen
  \bibfield  {author} {\bibinfo {author} {\bibfnamefont {M.}~\bibnamefont
  {Silva-Feaver}} \emph {et~al.},\ }\href {\doibase 10.1109/TASC.2016.2631425}
  {\bibfield  {journal} {\bibinfo  {journal} {IEEE Trans. Appl. Supercond.}\
  }\textbf {\bibinfo {volume} {27}},\ \bibinfo {pages} {1400204} (\bibinfo
  {year} {2017})},\ \Eprint {http://arxiv.org/abs/1610.09344} {arXiv:1610.09344
  [astro-ph.IM]} \BibitemShut {NoStop}%
\bibitem [{\citenamefont {Gramolin}\ \emph {et~al.}(2021)\citenamefont
  {Gramolin}, \citenamefont {Aybas}, \citenamefont {Johnson}, \citenamefont
  {Adam},\ and\ \citenamefont {Sushkov}}]{Gramolin:2020ict}%
  \BibitemOpen
  \bibfield  {author} {\bibinfo {author} {\bibfnamefont {A.~V.}\ \bibnamefont
  {Gramolin}}, \bibinfo {author} {\bibfnamefont {D.}~\bibnamefont {Aybas}},
  \bibinfo {author} {\bibfnamefont {D.}~\bibnamefont {Johnson}}, \bibinfo
  {author} {\bibfnamefont {J.}~\bibnamefont {Adam}}, \ and\ \bibinfo {author}
  {\bibfnamefont {A.~O.}\ \bibnamefont {Sushkov}},\ }\href {\doibase
  10.1038/s41567-020-1006-6} {\bibfield  {journal} {\bibinfo  {journal} {Nature
  Phys.}\ }\textbf {\bibinfo {volume} {17}},\ \bibinfo {pages} {79} (\bibinfo
  {year} {2021})},\ \Eprint {http://arxiv.org/abs/2003.03348} {arXiv:2003.03348
  [hep-ex]} \BibitemShut {NoStop}%
\bibitem [{\citenamefont {Brouwer}\ \emph {et~al.}(2022)\citenamefont {Brouwer}
  \emph {et~al.}}]{DMRadio:2022pkf}%
  \BibitemOpen
  \bibfield  {author} {\bibinfo {author} {\bibfnamefont {L.}~\bibnamefont
  {Brouwer}} \emph {et~al.} (\bibinfo {collaboration} {DMRadio}),\ }\href
  {\doibase 10.1103/PhysRevD.106.103008} {\bibfield  {journal} {\bibinfo
  {journal} {Phys. Rev. D}\ }\textbf {\bibinfo {volume} {106}},\ \bibinfo
  {pages} {103008} (\bibinfo {year} {2022})},\ \Eprint
  {http://arxiv.org/abs/2204.13781} {arXiv:2204.13781 [hep-ex]} \BibitemShut
  {NoStop}%
\bibitem [{\citenamefont {Berlin}\ \emph {et~al.}(2021)\citenamefont {Berlin},
  \citenamefont {D'Agnolo}, \citenamefont {Ellis},\ and\ \citenamefont
  {Zhou}}]{Berlin:2020vrk}%
  \BibitemOpen
  \bibfield  {author} {\bibinfo {author} {\bibfnamefont {A.}~\bibnamefont
  {Berlin}}, \bibinfo {author} {\bibfnamefont {R.~T.}\ \bibnamefont
  {D'Agnolo}}, \bibinfo {author} {\bibfnamefont {S.~A.~R.}\ \bibnamefont
  {Ellis}}, \ and\ \bibinfo {author} {\bibfnamefont {K.}~\bibnamefont {Zhou}},\
  }\href {\doibase 10.1103/PhysRevD.104.L111701} {\bibfield  {journal}
  {\bibinfo  {journal} {Phys. Rev. D}\ }\textbf {\bibinfo {volume} {104}},\
  \bibinfo {pages} {L111701} (\bibinfo {year} {2021})},\ \Eprint
  {http://arxiv.org/abs/2007.15656} {arXiv:2007.15656 [hep-ph]} \BibitemShut
  {NoStop}%
\end{thebibliography}%

\end{document}